# Quantum States Interrogation Using a Pre-shaped Free Electron Wavefunction

Bin Zhang[1], Du Ran[1,2,3,*], Reuven Ianconescu[1,4], Aharon Friedman[5], Jacob Scheuer[1], Amnon Yariv[6], Avraham Gover[1,†]

*Corresponding author: randu11111@163.com; †Corresponding author: gover@eng.tau.ac.il

[1] School of Electrical Engineering - Physical Electronics, Center of Laser-Matter Interaction, Tel Aviv University, Ramat Aviv 69978, Israel

[2] School of Electronic Information Engineering, Yangtze Normal University, Chongqing 408100, China

[3] Fujian Key Laboratory of Quantum Information and Quantum Optics, Fuzhou University, Fuzhou 350116, China

[4] Shenkar College of Engineering and Design 12, Anna Frank St., Ramat Gan, Israel

[5] Schlesinger Family Accelerator Centre, Ariel University, Ariel 40700, Israel

[6] California Institute of Technology (Caltech), Pasadena, California 91125, USA

**Abstract:** We present a comprehensive theory for interrogation of the quantum state of a two-level system (TLS) based on a free-electron – bound-electron resonant interaction scheme. The scheme is based on free electrons, whose quantum electron wavefunction is pre-shaped or optically modulated by lasers in an electron microscope setup and then inelastically scattered by a quantum TLS target (e.g., atom, quantum dot, crystal defect center, etc.) upon traversing in proximity to the target. Measurement of the post-interaction energy spectrum of the electrons, probes and quantifies the full Bloch sphere parameters of a pre-excited TLS and enables coherent control of the qubit states. The exceptional advantage of this scheme over laser-based ones, is atomic-scale spatial resolution of addressing individual TLS targets. Thus, this scheme opens new horizons for electron microscopy in material interrogation and quantum information technology.

## I. INTRODUCTION

Coherent control of a single quantum two-level system (TLS) using lasers is a subject of highest interest in quantum optics. The subject is central to understanding fundamental phenomena such as Rabi oscillation [1], Landau-Zener transition [2, 3] and Ramsey interference [4]. It also relates to coherent laser control of atoms [5]. Additionally, it is used extensively in quantum information technologies [6, 7] in connection with control and reading of quantum bits (qubits). A variety of TLS embodiments are considered in this context, including trapped ions [8], quantum dots [9] and defects in solids [10].

In this paper we propose an alternative scheme for interrogation and control of the quantum state of a TLS, with pre-shaped quantum electron wavepackets (QEW) rather than with a coherent laser beam. This scheme, based on the recently proposed concept of free-electron bound-electron resonant interaction (FEBERI) [11], differs from more conventional approaches by being able to address individual TLS targets with atomic-scale (nanometric) spatial resolution [12] in contrast to the micrometric resolution of a laser beam. This advantage stems from the fact that typical electron de-Broglie wavelength is six orders of magnitude shorter than that of the optical wavelength of a laser beam. Recent technological developments in electron-quantum-optics pave the way to implementing this alternative approach for coherent control and interrogation of TLS targets using pre-shaped and optically modulated QEWs. The diagnosis of the quantum state in this scheme relies on extension of well-established techniques in electron microscopy – Electron Energy Loss Spectroscopy (EELS) and cathodoluminescence (CL).

Recent technological advances enable the shaping of single QEWs in the transverse [13-15] and longitudinal [16-18] dimensions. Furthermore, it has been demonstrated that the QEW density expectation value can be modulated at optical frequencies by interaction with a laser beam [12, 19-29], utilizing the scheme of photo-induced near-field electron microscopy (PINEM) [30]. This modulation is detectable by interaction with a second, phase correlated, laser beam, attesting to the reality of the QEW periodic spatiotemporal sculpting (modulation) in the context of stimulated radiative interaction [20, 21, 31-33].

In considering the free electron wavefunction as an alternative to a laser beam for probing and controlling the TLS, a key question in the suggested new concept is the physical interpretation of the free electron wavefunction shape and the reality of the QEW shape in its interaction with matter. The reality of the QEW and the measurability of its shape and spatial dimensions, as well as the transition from the quantum wavefunction presentation to the classical point-particle theory (the wave-particle duality), have been considered previously in the context of electron interaction with light [34-42]. It was shown that the transition of the QEW radiative interaction from the wave-like regime, exhibiting characteristic multi-sidebands PINEM energy spectrum [19-31, 43, 44], to the point-particle-like acceleration/deceleration regime [45, 46], takes place when:

$$\Gamma = \omega \sigma_t < \sqrt{2} \tag{1}$$

Namely, the transition takes place when the wavepacket standard deviation duration $\sigma_t$ (or its longitudinal breadth $\sigma_z = v_0 \sigma_t$) is short relative to the optical radiation period $2\pi/\omega$ (or wavelength $\lambda$) [35-38, 46]. The reality interpretation of the QEW modulation features in the context of radiative emission, has been also extended to the all-important case of multiple modulation-correlated electron wavepackets, where coherent superradiant emission [47], proportional to the number of electrons squared $N^2$, is expected [38, 48-50], analogously the classical case of a pre-bunched point-particle beam [51].

In analogy with the interaction of a QEW with radiation, the reality of the QEW shaping and its modulation features were claimed to be manifested also in interaction with matter in a newly proposed effect of FEBERI [11]. Based on a simple semiclassical model, it was proposed in [11] that a QEW, passing in the vicinity of a TLS target (e.g., an atom, quantum dot, crystal color center, trapped ion), would induce transitions in the TLS which depend on the QEW profile. Specifically, it was suggested that an ensemble of optical frequency modulated QEWs would excite resonantly TLS transitions, if their modulation frequency (produced by a laser of frequency $\omega_b$ in a PINEM setup [19]) is a sub-harmonic of the TLS transition frequency:

$$n\omega_b = \omega_{2,1} = E_{2,1}/\hbar \tag{2}$$

where $E_{2,1} = E_2 - E_1$ is the energy separation of the TLS quantum levels. In addition, it was argued, that if all QEWs are modulation-correlated (i.e. modulated by the same coherent laser), the transition rate should be enhanced in proportion to $N^2$, in analogy with the superradiance effect of an ensemble of atoms [47] or a bunched point-particles electron beam [31, 51]. Quadratic dependence of coherently modulated electron beam scattering efficiency by bound electron quantum states has been recognized earlier also as "pulsed beam scattering" [50, 52, 53]. In the case of a beam of multiple near-point-particle QEWs it was recently termed "Quantum Klystron" [54] in connection to microwave frequency quantum transition. When such near-point-particle QEWs are injected as a periodic train of pulses they would produce FEBERI transition rate enhanced in proportion to $N^2$.

The semiclassical analysis of FEBERI, based on Born's probability interpretation of the wavefunction envelope modulation [11, 55], and its proposed dependence on the QEW dimensions, were questioned [56, 57]. This led to a furry of recent publications on the subject [23, 24, 54-62]. Nevertheless, recent, fully quantum-mechanical analyses (of both the free and bound electrons), have substantiated the proposition of shape- and modulation-dependence of multiple correlated QEWs interactions with a TLS and with radiation [48, 55]. Moreover, technological and conceptual developments of schemes for optical control of the temporal shape of the electron wavefunction, show that attosecond-scale pre-shaping and modulation of QEWs are experimentally realizable [16, 17], and can be used to realize the proposed theoretical concept.

Based on these conceptual and technological developments, we propose here that the shape-dependence of the QEW interaction with a TLS can be utilized for interrogation of the quantum excitation state of a TLS [48, 55, 58, 61, 63]. We present a comprehensive theory of quantum state interrogation by FEBERI, showing that the initial quantum state (qubit) coordinates on the Bloch sphere, which uniquely define superposition TLS quantum states (TLS-QS), can be fully extracted from the post-interaction electron energy spectrum of the interacting QEWs, and coherently controlled by their pre-shaping. This can be achieved by controlling the size of the QEWs before interaction, or pre-modulating them at optical frequency by a PINEM process. Interrogation of the TLS state has been analyzed in [58] considering an energy modulated plane-wave quantum electron wavefunction. In [61], a particular energy distribution (coherent superposition state) of the free electron wavefunction was considered and shown to be useful for evaluating the decoherence and relaxation parameters of an excited TLS. The present work depicts a more comprehensive quantum model in which the free electron wavefunction is described in terms of a QEW of general distribution in the momentum or spatial domains. The use of both presentations reveals that the entanglement of the free and bound electrons results in an energy spectrum of the post-interaction QEWs that corresponds to (and thus interrogates) the initial TLS-QS and depends on the spatiotemporal shape and size of the QEWs (the wavepacket envelope size). We show in this paper that a broad size QEW (nearly plane-wave quantum wavefunction) carries the fingerprint of the polar angle of the qubit state of a target TLS on the Bloch sphere, with little energy transfer, and therefore can be useful for its interrogation. On the other hand, interaction with narrow finite size QEWs (nearly point-particle limit) or optical-frequency density modulated QEWs, involves energy transfer between the free electron and the TLS, and enable also its coherent control.

The analysis here refers to the probability of quantum electron transitions of a single TLS and the corresponding modification of the energy density spectrum of an interacting single QEW. Obviously, measurement of the free electron energy probability distribution requires multiple QEWs and (or) multiple TLS targets. Fundamentally, inelastic scattering of an electron that results in quantum transition of a TLS-QS to the upper or lower quantum level, destroys the state, and therefore interrogation of a single TLS-QS by measuring the post-interaction EELS of multiple electrons, is possible only if the TLS-QS is re-instated after each interaction. We conjecture that coherent control and interrogation of the TLS state with the joint quantum wavefunction of multiple modulation-correlated electrons can be used for control and interrogation of the TLS state [63]. This conjecture is an extension of the recent conceptual proposals of superradiance [38, 48, 49] and Rabi-oscillation FEBERI [11, 55] processes with multiple optical-frequency modulation-correlated electrons. However, the analysis here is confined to interaction of single pre-shaped and pre-modulated QEWs with a single TLS target. The extension to multiple particles [11, 48, 63] and targets [61, 64] is commented upon in the discussion part.

Based on this new theory and formulation, we expect that further development of the proposed concept would lead to implementation of the FEBERI effect in numerous important applications of electron microscopy in atomic-scale probing of quantum excitations in matter, diagnostics and coherent control of qubits [65] and efficient pumping of single quantum dot quantum emitters [66].

## II. THEORETICAL FRAMEWORK

We present analysis of a general experimental scheme for interrogating bound-electron quantum states by free electrons. The free electron is modeled as a quantum electron wavepacket (QEW) propagating in proximity to a Hydrogen-like atom that models a general quantum two-level system (TLS). We assume an experimental setup, as shown in Fig. 1, in which the TLS quantum state is pre-set by a coherent laser beam that is phase locked to a laser beam that pre-shapes or pre-modulates the QEW with controlled delay. We assert that the quantum state of the TLS can be deduced from the post-interaction electron energy spectrum of the free electrons which is accumulated by repeated reinjection of identical QEWs and re-excitation of the TLS.

We start by describing the TLS as an electric dipole [55], and assuming that the interaction between the free electron and the TLS is dominated by Coulomb potential. The joint wavefunction of the free and bound electrons is governed by the relativistically modified Schrodinger equation:

$$i\hbar \frac{\partial}{\partial t}|\Psi\rangle = (\widehat{H}_0 + \widehat{H}_I)|\Psi\rangle \tag{3}$$

where $\widehat{H}_0 = \widehat{H}_{0F} + \widehat{H}_{0B}$, with $\widehat{H}_{0F}, \widehat{H}_{0B}$ being the kinetic Hamiltonian of the free electron and the Hamiltonian of the bound electron respectively, and $\widehat{H}_I$ is the interaction Hamiltonian. For the analysis to be valid also for relativistic electrons, we use the relativistic 1D kinetic Hamiltonian for a free electron of energy $E_0 = \gamma m_e c^2$ and momentum $p_0 = \gamma m_e v_0$, where $m_e = 9.1 \times 10^{-31} kg$ is the electron mass, $v_0$ is the velocity of the free electrons ($\beta = v_0/c = 0.7$ in our work, $c$ is the velocity of light):

$$\widehat{H}_{0F} = E_0 + v_0 \cdot (\hat{p} - p_0) + \frac{1}{2\gamma^3 m_e}(\hat{p} - p_0)^2 \tag{4}$$

This Hamiltonian was derived in Refs. [35, 36] by a second order iterative approximation of Klein-Gordon equation, neglecting the spin effects. The Hamiltonian satisfies $\widehat{H}_{0F}|p\rangle = E_p|p\rangle$, where $|p\rangle$ is the eigenstate of momentum operator $\hat{p}$ which satisfies $\langle p|p'\rangle = \delta(p-p')$ and $\int dp\, |p\rangle\langle p| = I$, and the energy dispersion relation is $E_p = E_0 + v_0(p - p_0) + \frac{1}{2\gamma^3 m_e}(p - p_0)^2$. The TLS Hamiltonian satisfies $\widehat{H}_{0B}|i\rangle = E_i|i\rangle$ (i = 1,2), where $E_i = \hbar\omega_i$ and $|i\rangle$ is the eigenstate of the TLS which satisfies $\langle i|j\rangle = \delta_{ij}$.

The neglect of exchange energy and the spin-orbit interaction effects is valid in our simplified model under the assumption that the free and bound electrons do not overlap spatially. There is no interaction between the pre-shaped QEW and the laser field used to excite the TLS assuming that it is vanished by the time the QEW meets the TLS. Thus, the dominant interaction in the system is the Coulomb interaction of the dipole and free electron, yielding an interaction Hamiltonian $\widehat{H}_I = -e\hat{r}' \cdot \boldsymbol{E}(\hat{z}, r_{\perp 0})$, where $\hat{r}'$ is the position vector of the bound electron relative to the model Hydrogen nucleus, and $\hat{r} = (\hat{z}, \boldsymbol{r}_{\perp 0})$ is the position vector of the free electron relative to the nucleus (See Fig. 1). In a 1D model, the free electron is presented as a QEW of negligible narrow width ($\boldsymbol{r} = (z, 0)$). Such a relativistic electron generates an electric field $\boldsymbol{E}(z, \boldsymbol{r}_{\perp 0})$ at the TLS position is [67]:

$$\boldsymbol{E}(z, \mathrm{r}_{\perp 0}) = -\frac{e\gamma}{4\pi\epsilon_0} \frac{z\hat{\boldsymbol{e}}_z + r_{\perp 0}\hat{\boldsymbol{e}}_\perp}{(\gamma^2 z^2 + r_{\perp 0}^2)^{\frac{3}{2}}} \qquad (5)$$

where $\hat{\boldsymbol{e}}_z$ and $\hat{\boldsymbol{e}}_\perp$ represent the directions along and perpendicular to the free electron propagation, respectively.

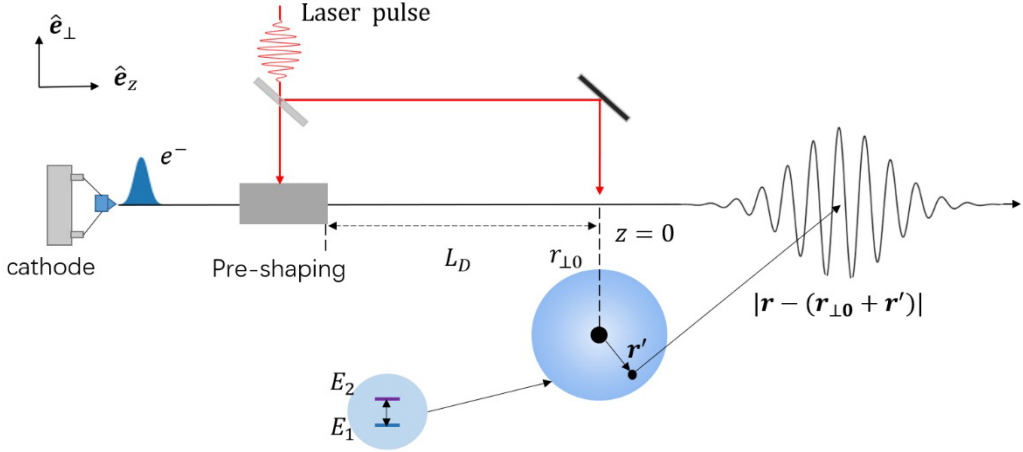

FIG. 1. Scheme of TLS interrogation with FEBERI. The free electron QEW is emitted from e-the gun and after acceleration gets pre-shaped or energy-modulated through interaction with the laser-induced optical near-field of a tip/grating/membrane/nano-resonator (a PINEM process). After traversing a free-drift length $L_D$ along the z axis, the shaped or density-modulated QEW interacts with a bound electron at coordinate (z=0, $\boldsymbol{r}_{\perp 0} + \boldsymbol{r}'$). The TLS is pre-excited into a quantum superposition state by a phase-locked harmonic of the same laser source with a controlled delay. The FEBERI interaction can be measured through the energy spectrum of the free electron.

We take the initial state of the free electron to be a general quantum state: $|\psi_F(p,t)\rangle = \sum_p c_p^{(0)} e^{-iE_p t/\hbar} |p\rangle$, where $c_p^{(0)}$ is the quantum electron wavefunction in momentum space. The general time-dependent pre-interaction state of the bound electron is represented as $|\psi_B(t)\rangle = \sin\left(\frac{\theta}{2}\right) e^{-i\omega_1 t} |1\rangle + e^{i\phi} \cos\left(\frac{\theta}{2}\right) e^{-i\omega_2 t} |2\rangle$, where $\theta \in [0, \pi]$ is the polar angle and $\phi \in [0, 2\pi]$ is the azimuth angle of the Bloch sphere [68]. The combined wavefunction of the free and bound electrons $|\Psi\rangle$, is the solution of Eq. 1 in the presence of the electric dipole interaction which entangles the free and bound electron states.

It is convenient to transform to the interaction picture by using the unitary evolution operator $\widehat{U}(t) = \exp[-i\widehat{H}_0 t/\hbar]$ such that Eq. 3 becomes

$$i\hbar \frac{\partial}{\partial t} |\widetilde{\Psi}(p,t)\rangle = \widehat{\mathcal{H}}_I(t) |\widetilde{\Psi}(p,t)\rangle \qquad (6)$$

where $|\widetilde{\Psi}(p,t)\rangle = \widehat{U}^\dagger(t) |\Psi(p,t)\rangle$, and the Hamiltonian $\widehat{\mathcal{H}}_I(t) = \widehat{U}^\dagger(t) \widehat{H}_I \widehat{U}(t) = \boldsymbol{\mu}_{2,1} \cdot \boldsymbol{E}(\hat{z} + v_0 t, r_\perp) e^{-i\omega_{2,1} t} \hat{\sigma}^+ + h.c.$ with $\boldsymbol{\mu}_{2,1} = -e\langle 2|\hat{r}'|1\rangle$ being the dipole moment of the TLS, and $\hat{\sigma}^+ = |2\rangle\langle 1|$ ($\hat{\sigma}^- = |1\rangle\langle 2|$) being the TLS raising (lowering) operator, see Appendix A for the detailed derivation. Thus, the wavefuction after interaction is obtained by multiplying the initial wavefunction with the scattering operator $\hat{S}(t_f, t_i) = \mathcal{T} \exp\left[-\frac{i}{\hbar} \int_{t_i}^{t_f} \widehat{\mathcal{H}}_I(t) dt\right]$, where $\mathcal{T}$ is the time-ordering operator, $t_i$ and $t_f$ are the initial time and final times of the interaction (see Appendix B):

$$|\widetilde{\Psi}_f(p, t_f)\rangle = \hat{S}(t_f, t_i)|\widetilde{\Psi}_i(p, t_i)\rangle \tag{7}$$

Since the coupling strength between free electron and bound electron is relatively weak, it is reasonable to consider only the first order expansion of the scattering matrix, which is equivalent to first order perturbation theory. Then (derived in Appendix B):

$$\hat{S}(t_f, t_i) = \cos|g| - i \sin|g| \left( e^{i\phi_g}\hat{\sigma}^+ e^{-i\frac{\omega_{2,1}}{v_0}\hat{z}} + h.c. \right) \tag{8}$$

where $g = \frac{\boldsymbol{\mu}_{2,1}}{\hbar v_0} \cdot \int_{-\infty}^{\infty} du\, \boldsymbol{E}(u, r_{\perp 0}) e^{-i\omega_{2,1}u/v_0}$ is the coupling coefficient between the free electron and the TLS in the limit $t_i \to -\infty, t_f \to \infty$ [55]:

$$g = -\frac{e}{2\pi\epsilon_0\gamma\hbar}\frac{\omega_{2,1}}{v_0^2}\boldsymbol{\mu}_{2,1} \cdot \left[\frac{i}{\gamma}K_0\left(\frac{\omega_{2,1}r_\perp}{v_0\gamma}\right)\hat{\boldsymbol{e}}_z + K_1\left(\frac{\omega_{2,1}r_\perp}{v_0\gamma}\right)\hat{\boldsymbol{e}}_\perp\right] = |g|e^{i\phi_g} \tag{9}$$

For any input state $|\widetilde{\Psi}_i\rangle = |\widetilde{\psi}_F^i\rangle \otimes |\widetilde{\psi}_B^i\rangle$, we can get a general form of the final state $|\widetilde{\Psi}_f\rangle$ by using the scattering operator (Eq. 8) in Eq. 7, where the general form of the initial TLS state and QEW in the interaction picture is represented by $|\widetilde{\psi}_B^i\rangle = \sin\frac{\theta}{2}|1\rangle + e^{i\phi}\cos\frac{\theta}{2}|2\rangle$ and $|\widetilde{\psi}_F^i(p)\rangle = \sum_p c_p^{(0)}|p\rangle$ respectively. The following derivation step (see Appendix C) is a calculation of the post-interaction density matrix $\hat{\rho}_f = |\widetilde{\Psi}_f\rangle\langle\widetilde{\Psi}_f|$ of the quantum-entangled free and bound electron systems. Finally, the density matrices of the bound and free electrons are derived by partial tracing: $\hat{\rho}_F^f = \text{Tr}_B[\hat{\rho}_f]$ and $\hat{\rho}_B^f = \text{Tr}_F[\hat{\rho}_f]$, respectively. This results in the explicit expressions for the free electron post-interaction momentum density distribution for any initial electron wavefunction $c_p^{(0)}$:

$$\begin{aligned}\rho^f(p) = \text{diag}(\hat{\rho}_F^f) &= \cos^2|g|\left|c_p^{(0)}\right|^2 + \sin^2|g|\left[\cos^2\left(\frac{\theta}{2}\right)\left|c_{p+\delta p_{2,1}}^{(0)}\right|^2 + \sin^2\left(\frac{\theta}{2}\right)\left|c_{p-\delta p_{2,1}}^{(0)}\right|^2\right] \\ &- \frac{1}{2}\sin(2|g|)\sin\theta\, \text{Re}\left\{ie^{-i(\phi-\phi_g)}Co(p, \delta p_{2,1}) + ie^{i(\phi-\phi_g)}Co(p, -\delta p_{2,1})\right\}\end{aligned} \tag{10}$$

where $\delta p_{2,1} = \hbar\omega_{2,1}/v_0$ is the electron recoil parameter and $Co(p, \delta p) \equiv c_{p+\delta p}^{*(0)}c_p^{(0)}$.

Complementary to the derivation of the QEW momentum distribution, the corresponding final occupation probabilities of the upper and lower levels of the bound electron are (see Appendix C):

$$\begin{aligned}P_1^f &= \langle 1|\hat{\rho}_B^f|1\rangle = P_1^i + \sin^2|g|\cos\theta - \frac{1}{2}\sin(2|g|)\sin\theta\, \text{Re}\left\{I(-\delta p_{2,1})e^{-i(\phi_g-\phi)+i\frac{\pi}{2}}\right\} \\ P_2^f &= \langle 2|\hat{\rho}_B^f|2\rangle = P_2^i - \sin^2|g|\cos\theta - \frac{1}{2}\sin(2|g|)\sin\theta\, \text{Re}\left\{I(\delta p_{2,1})e^{i(\phi_g-\phi)+i\frac{\pi}{2}}\right\}\end{aligned} \tag{11}$$

where $I(\delta p_{2,1}) = \int_p dp\, Co(p, \delta p_{2,1}) = \int_p dp\, \langle c_{p+\delta p_{2,1}}^{(0)}|c_p^{(0)}\rangle$ is an auto-correlation function [55, 58].

Eq. (11) satisfies $P_1^f + P_2^f = 1$. Furthermore, Eq. (10) and Eq. (11) result in together a conservation of energy relation between the free electron and bound electron subsystems:

$$E_{2,1} \cdot \Delta P_2 = \int E_p \rho^f(p) dp - E_0 \tag{12}$$

where $\Delta P_2 = P_2^f - P_2^i = -\Delta P_1$. This manifests the self-consistence of the presented formulation.

## III. INTERROGATION OF A TLS WITH SHAPE-CONTROLLED QEWS

In this section, we analyze the interaction between a TLS and a finite-size QEW, and present the dependence of the post-interaction incremental spectrum of the free electron on the QEW size. We demonstrate the wave-particle duality of the QEW in interaction with matter and its transition from quantum plane-wave to near-point-particle characteristics. This transition is analogous to the wave-particle duality and the transition from multi-photon emission/absorption (PINEM) process to acceleration/deceleration process in interaction of QEWs with light that we have analyzed earlier [36, 37, 38].

We model the free electron as a general Gaussian wavepacket including broadening and momentum chirp due to energy-dispersive drift from a controlled waist point located a distance $L_D = v_0 t_D$ before the interaction point $z = 0$ [36] (see Appendix D). In momentum representation:

$$c_p(t_D) = \frac{1}{(2\pi\sigma_p^2)^{\frac{1}{4}}} e^{-\frac{iE_p t_D}{\hbar}} \exp\left[-\frac{(p-p_0)^2}{4\sigma_p^2}\right] = \frac{e^{-\frac{i[E_0 + v_0(p-p_0)]t_D}{\hbar}}}{(2\pi\sigma_p^2)^{\frac{1}{4}}} \exp\left[-\frac{(p-p_0)^2}{4\tilde{\sigma}_p^2(L_D)}\right] \tag{13}$$

where $\sigma_p$ is the wavepacket momentum spread. We lumped the quadratic term of Eq. (4) into the definition of a complex momentum spread parameter $\tilde{\sigma}_p(L_D) = \sqrt{\frac{\sigma_p^2}{1 + i\frac{2L_D}{\gamma^3 m v_0 \hbar} \sigma_p^2}}$. Correspondingly, the QEW in space-time coordinates is:

$$\psi_z(L_D) = \frac{1}{(2\pi\tilde{\sigma}_z^2(L_D))^{\frac{1}{4}}} \exp\left[-\frac{(z - v_0 t_D)^2}{4\tilde{\sigma}_z^2(L_D)}\right] \tag{14}$$

where $\tilde{\sigma}_z(L_D) = \sigma_{z0}\sqrt{1 + i\frac{\hbar L_D}{2\gamma^3 m_e v_0 \sigma_{z0}^2}}$, $\sigma_{z0} = \frac{\hbar}{2\sigma_p}$ is the QEW longitudinal waist size. The Gaussian wavepacket in real space and time coordinates $\sigma_t(t_D) = |\tilde{\sigma}_z(L_D)|/v_0$ broadens upon propagation from the location of its waist to the interaction point $z = 0$ after a drift time $t_D = L_D/v_0$. This history-dependent size of the QEW can be written as [36]:

$$\sigma_t(L_D) = \sigma_{t0}\sqrt{1 + L_D^2/z_{R\parallel}^2} \tag{15}$$

where $\sigma_{t0} = \sigma_t(L_D = 0)$ and we define a "longitudinal Rayleigh length" parameter $z_{R\parallel} = 4\pi\gamma^3\beta\sigma_{z0}^2/\lambda_c$, analogous to transverse expansion Rayleigh length of a laser beam, corresponding to the distance where the QEW envelope broadens by a factor of $\sqrt{2}$, $\lambda_c = h/m_e c$ is the Compton wavelength.

Using Eq. (13) as the initial state $c_p^{(0)}$ in Eq. (10), we get the post-interaction momentum density distribution of the Gaussian QEW (details shown in Appendix E):

$$\rho_p^f = \rho_p^i + \Delta\rho = \rho_p^i + \Delta\rho_p^{(0)} + \Delta\rho_p^{(1)} + \Delta\rho_p^{(2)} \tag{16}$$

where $\rho_p^i = |c_p(t_D)|^2 = \frac{1}{\sqrt{2\pi\sigma_p^2}} exp\left[-\frac{(p-p_0)^2}{2\sigma_p^2}\right]$

$$\Delta\rho_p^{(0)} = -\sin^2|g| \frac{1}{\sqrt{2\pi\sigma_p^2}} \exp\left[-\frac{(p-p_0)^2}{2\sigma_p^2}\right] \tag{17}$$

$$\Delta\rho_p^{(1)} = \frac{\sin(2|g|)}{2} \sin\theta \, e^{-\frac{\Gamma^2}{2}} \sin(\zeta + \phi_g) \frac{1}{\sqrt{2\pi\sigma_p^2}} \left\{ e^{-\frac{\left(p-p_0+\frac{\delta p_{2,1}}{2}\right)^2}{2\sigma_p^2}} - e^{-\frac{\left(p-p_0-\frac{\delta p_{2,1}}{2}\right)^2}{2\sigma_p^2}} \right\} \tag{18}$$

$$\Delta\rho_p^{(2)} = \sin^2|g| \frac{1}{\sqrt{2\pi\sigma_p^2}} \left[ \cos^2\left(\frac{\theta}{2}\right) e^{-\frac{(p-p_0+\delta p_{2,1})^2}{2\sigma_p^2}} + \sin^2\left(\frac{\theta}{2}\right) e^{-\frac{(p-p_0-\delta p_{2,1})^2}{2\sigma_p^2}} \right] \tag{19}$$

The corresponding increments of the occupation probabilities of the TLS quantum levels that are probed by this incremental energy spectrum are (from Eq. 11):

$$\Delta P_2 = -\Delta P_1 = -\sin^2|g|\cos\theta + \frac{1}{2}\sin(2|g|)\sin\theta\sin(\zeta + \phi_g)\, e^{-\frac{(\Gamma^2+\Gamma_D^2)}{2}} \tag{20}$$

where $\Gamma = \omega_{2,1}\sigma_t(L_D)$ and $\Gamma_D = \frac{1}{2\omega_{2,1}\sigma_{t0}}\frac{L_D}{z_G} = \frac{\sigma_E}{E_{2,1}}\frac{L_D}{z_G}$, with [36]

$$z_G = 2\gamma^3\beta^3 \frac{m_e c^3}{\hbar\omega_{2,1}^2} \tag{21}$$

Note that the decay constant $\Gamma$ is the same as the decay constant (Eq. 1) of radiative interaction of a QEW in transition from wave to point-particle limit [36, 69] with the substitution $\omega = \omega_{2,1}$. So, to avoid diminishing of the first order FEBERI interaction, one should keep $\Gamma = \omega_{2,1}\sigma_t(L_D) < 1$ and $L_D < z_G$ (see Appendix D).

The phase $\phi$ is the phase of the superposition quantum state of the TLS (azimuthal angle of the qubit) at the time of the TLS excitation (see Fig. 1), and $\zeta = \omega_{2,1}\Delta t - \phi$ is the phase at the arrival time of the QEW centroid to the TLS location $z = 0$. $\Delta t$ is the time elapsed from the TLS excitation to the arrival of the QEW. It is the difference between the electron drift time $t_D$ and the optical path delay time between the TLS excitation and the QEW formation by the phase-locked laser beams (see Fig. 1).

Figure 2 demonstrates the dependence of the free electron post-interaction energy spectrum on the parameters of the TLS-QS on the Bloch sphere $(\theta, \phi)$. For simplicity we consider here the case when the QEW arrives at the interaction point at its minimal waist without chirp, namely, $\sigma_t(L_D = 0) = \sigma_{t0} = \hbar/2\sigma_E$. Figures (2a), (2d) and (2e) display the incremental energy density distribution $\Delta\rho_{E_p}$ of the free electron after interaction that was calculated numerically (see Appendix H) for different wavepacket sizes $\sigma_t$. For comparison we overlay in Fig. (2e) the analytically calculated curves of Eqs. (16-19), showing excellent agreement. The curves were calculated for parameters $g = 1.1 \times 10^{-3}, r_{\perp 0} = 2\, nm, \beta = 0.7\, (\gamma = 1.4)$ corresponds to the free electron energy $205\, keV$, the TLS transition energy $E_{2,1} = 1.97\, eV$, $\omega_{2,1} = 3\, fs^{-1}$ and period $T_{2,1} =$ $2\pi/\omega_{2,1} = 2.1\, fs$. Animated display of these dependences in the continuous parameters range ($0 \leq \theta \leq \pi, 0 \leq \phi \leq 2\pi$) is given in movies S1-S3.

Figure (2d) displays the dependence of the incremental energy spectrum on the polar parameter $\theta$ of the TLS state under the condition of large quantum recoil (narrow energy spread):

$$\delta p_{2,1} = \frac{\hbar \omega_{2,1}}{v_0} > 2\sigma_p \quad (E_{2,1} > 2\sigma_E) \tag{22}$$

In this limit, the spectrum displays a PINEM-like spectrum of well distinguished two sidebands peaked at $E_P - E_0 = \pm E_{2,1}$. The asymmetry between the negative and positive recoil sidebands reflects the different dependencies on the polar angle $\theta$ of the two sideband terms in the second order increment term $\Delta \rho_p^{(2)}$ (Eq. 19), which is dominant in this limit. In general, the first order term $\Delta \rho_p^{(1)}$ (Eq. 18) would add to the spectrum two more azimuthal angle dependent anti-symmetric sidebands peaked at $\pm E_{2,1}/2$, but these do not show up in the case depicted in Fig. (2d) since it is diminished by the factor $\exp(-\Gamma^2/2)$. This decay factor is dominant in the case of long QEW shown in Fig. (2d), because under condition (Eq. 22) $\Gamma = \omega_{2,1}\sigma_t = E_{2,1}/\sigma_E > 1$, and the first order term decays then exponentially below the second order term. However, this is not necessarily the case when $\sigma_t$ is not very large. Since the first and second order terms are proportional respectively to $|g|$ and $|g|^2$, the rigorous condition for neglecting the contribution of the first order term depends on $|g|$, namely, for $|g| \ll 1$ it is $\Gamma = \omega_{2,1}\sigma_t > \sqrt{-2\ln|g|}$. It is instructive to note that unlike the spectrum of PINEM [19], the post-interaction spectrum in Fig. (2d) displays only two first-order sidebands, which implies that only one quantum of energy is exchanged between the free electron and the TLS in the FEBREI process for a single electron interaction.

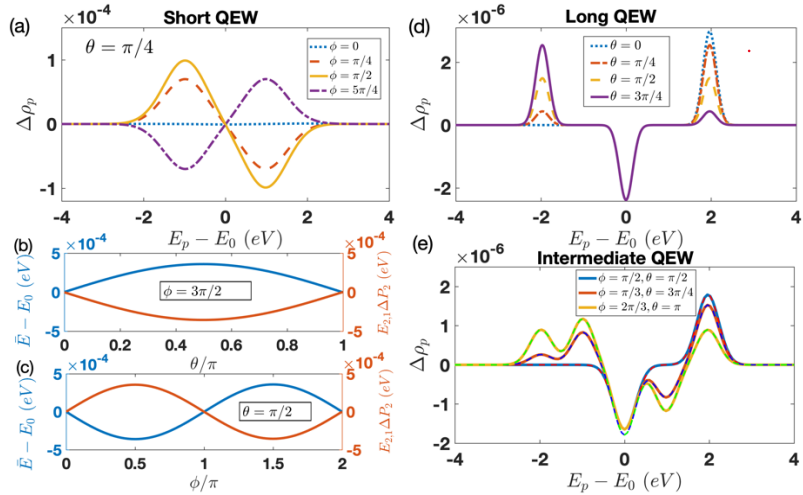

FIG. 2. Incremental energy spectra for different sizes of QEWs. (a) For a short QEW ($\sigma_{t0} = 0.3T_{2,1}$), $\Delta \rho_p$ is dominated by the first-order term Eq. (18) and exhibits acceleration/deceleration spectrum dependent on the azimuthal phase $\phi$ (shown for the latitude angle $\theta = \pi/4$). (b) and (c) show the incremental energies of the QEW (left-side axis) and the TLS (right-side axis) as a function of $\theta$ and $\phi$ respectively, demonstrating the energy conservation relation (Eq. 12). (d) For a long QEW ($\sigma_{t0} = T_{2,1}$), the magnitude of sidebands is determined by the polar angle $\theta$ independently of $\phi$. (e) For an intermediate case QEW ($\sigma_{t0} = 0.6T_{2,1}$), the spectrum provides information on both polar ($\theta$) and azimuth ($\phi$) angles (see video displays of continuous angle dependencies in supplemental material movies S1-S3 [70]). The dashed curves in Fig. 2 (e) display the results of the analytical formulae Eqs. (16-19) and match the numerical results (continuous curves) very well.

Figure (2a) displays the dependence of the incremental energy spectrum on the azimuthal parameter $\phi$ of the TLS state under the condition of short (near point particle) QEW:

$$2\sigma_t(L_D) < T_{2,1} \tag{23}$$

In this limit, the curves display an acceleration/deceleration spectrum that depends according to Eq. (18) sinusoidally on the polar $\theta$ and azimuthal $\phi$ angles of the TLS-QS (shown in Fig. (2a) for $\theta = \pi/2$). In this case $\Gamma = \omega_{2,1}\sigma_t < 1$ and the decay factor is $e^{-\Gamma^2/2} \approx 1$, thus the linear term $\Delta\rho^{(1)}$ is dominant. The opposite polarity positive and negative recoil terms in Eq. 18 (peaked at $\pm E_{2,1}/2$) do not show up in Fig. (2a), because in this case $2\sigma_E > E_{2,1}$. Consequently, these terms overlap, and the incremental energy distribution function displays an S-shaped curve, reflecting through $\zeta$ the phase $\phi$ of the TLS quantum state at the arrival time of the QEW centroid. Figures (2b) and (2c) display the total energy expectation value increment of the post-interaction QEW as function of the polar ($\theta$) and azimuth ($\phi$) angles of the TLS-QS. They also show the corresponding change of the TLS excitation incremental energy expectation value, in confirmation of the conservation of energy relation (Eq. 12). Figure (2e) displays an intermediate parameters case $2\sigma_t \simeq T_{2,1}$ in which the first (Eq. 18) and second order (Eq. 19) terms have similar weight, and the incremental energy spectrum reflects at the same time both the polar and azimuthal angles of the TLS state.

**Quantum wave-particle duality and anomalous FEBERI**

The interpretation of the quantum electron wavefunction $\Psi(\mathbf{r}, t)$ has been a matter of debate since the inception of quantum theory [71, 72]. The accepted Born interpretation is that the expectation value of the electron wavefunction modulus in space-time - $|\Psi(\mathbf{r}, t)|^2$, represents the probability density of finding the electron at point $\mathbf{r}$ at time t. While we use both momentum and space-time representations of the electron wavefunction, the latter is most conducive to physical understanding of the electron particle-like and wave-like regimes of interaction with light and matter in view of Born's interpretation of the wavefunction. As shown in Eq. (22), a narrow QEW in momentum (or energy) presentation interacts with a TLS as a quantum wave, exhibiting positive and negative recoils. On the other hand, the QEW interacts as a particle in the limit of a short wavepacket in space-time presentation (Eq. 23). In this case, consistently with Born's picture, the QEW is a near-point particle, experiencing acceleration or deceleration depending on its arrival time relative to the dipole moment oscillation phase of the TLS.

From the analysis of the previous section, we conclude that the interrogation of the quantum state of a pre-set TLS superposition state is possible by measuring the post-interaction energy spectrum of a passing-by QEW, as shown in Fig. 1. For probing the qubit polar angle $\theta$ of the TLS-QS it is necessary to employ a narrow energy spread QEW that satisfies the *large recoil condition* (Eq. 22). In this case the QEW is broad in real space, and its post-interaction spectrum exhibits a wave-like kind PINEM sidebands spectrum corresponding to positive and negative electron quantum recoils (Fig. 2d). The asymmetry of the sidebands pattern measures the polar angle $\vartheta$ of the TLS state. To probe the azimuth angle $\phi$ of the TLS quantum state, it is necessary to employ a short wavepacket that satisfies the near point-particle condition (Eq. 23), exhibiting an acceleration/deceleration post-interaction energy spectrum (Fig. 2a). However, the two conditions of large recoil (Eq. 22) and short wavepacket (Eq. 23) are complementary only in the case of a QEW at its waist ($t_D = 0$ in Eq. 14). In the general case $t_D \neq 0$, one must consider that the general QEW (Eq. 14) describes a chirped Gaussian with a longitudinal waist size $\sigma_t(t_D)$ broadened relative to the waist size $\sigma_{t0}$. The waist is located at an arbitrary waist location that can be controlled in principle at the QEW pre-shaping section of the experimental setup (Fig. 1) by optical shaping and chirping techniques [16, 17].

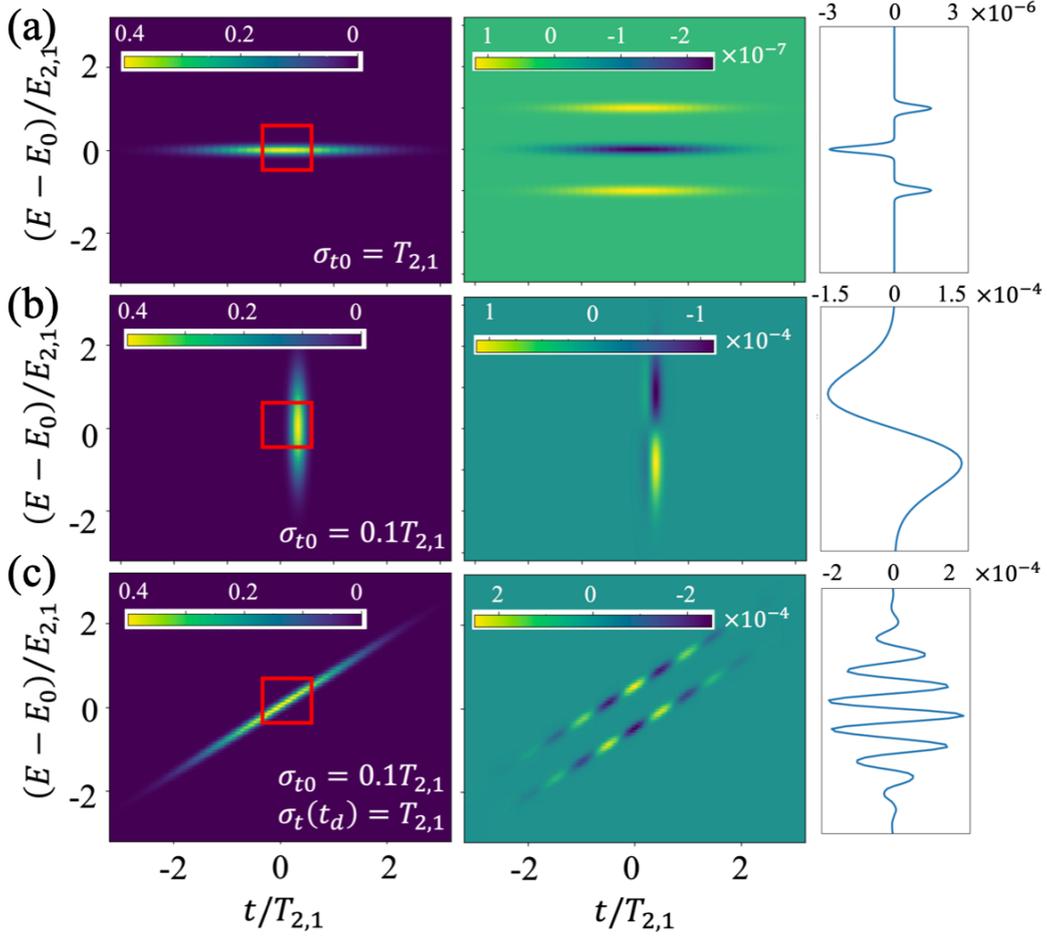

FIG. 3. Energy-time phase-space presentation and post-interaction incremental energy spectrum of the QEW. The first column displays the Wigner distribution of the QEW before interaction, and the second column displays the incremental Wigner distribution after interaction. The area of the pre-interaction Wigner distributions in the first column is the Planck constant $h$, the same as the square that represents the phase space area of the TLS $E_{2,1} \times T_{2,1}$. Consequently, the phase-space topography of the initial phase-space distribution of the QEW (left column) delineates only three possible cases, corresponding to: (a) Long QEW with large recoil, (b) Short QEW and (c) Chirped QEW. The second column depicts the incremental Wigner distribution of the QEW after interaction for TLS-QS parameters $\theta = \pi/2$, $\phi = \pi/2$. The third column presents the incremental energy density distribution before and after interaction, i.e., the difference of the horizontal projections of the Wigner distributions in the second and first columns.

In the general case the QEW may be chirped. The best way to describe this case is in phase space. Fig. 3 depicts the different regimes of QEW interaction with a TLS in energy-time (or momentum-space) phase-space. The Wigner distribution of a general chirped QEW (Eq. 13 or Eq. 14) after drift, can be calculated numerically from the post-interaction density matrix of the free electron (Eq. C5 in the Appendix). The pre-interaction distribution of the QEW upon entrance to the FEBERI interaction point, is shown in color code in the left column of Fig. 3 for three cases. Cases A and B correspond to injection of the QEW at its waist (no chirp) at the large recoil (Eq. 22) and narrow wavepacket (Eq. 23) conditions respectively. Similar to the representation of anomalous PINEM in Ref. [37], we overlay the Wigner distribution of the pre-interaction QEW on a phase-space square of area $E_{2,1} \times T_{2,1}$ representing the TLS, both have the area $h$ - the Planck constant. The second column in Fig. 3 depicts the incremental Wigner distribution after interaction, and the third column presents the incremental energy distribution density $\Delta\rho_p$, which is the difference of the horizontal projections of the Wigner distributions before and after interaction. We observe that the topology of the phase-space diagrams reveals a QEW parameters regime where neither the large recoil (Eq. 22) nor the narrow wavepacket (Eq. 23) conditions are satisfied. This regime is depicted in row C of Fig. 3. This regime is analogous to "anomalous PINEM" in the general case of PINEM interaction [37]. Evidently, one should avoid this chirp-dominated topological "no-man land" in phase-space, which does not reflect the TLS-QS qubit coordinates. This can be done by optical control of the wavepacket size and chirp as shown in [16, 17].

The control of the size and chirp of the QEW requires advanced optical techniques. In particular, the azimuthal phase interrogation scheme requires sub-optical-cycle shape control. Substantial progress has been achieved in QEW shape control by THz and IR chirp techniques [16, 17] and other schemes [18]. Assuming such control is available, the QEW breadth, chirp and waist-location parameters can be controlled at the QEW pre-shaping box (Fig. 1). This control is necessary if one wants to operate at the short wavepacket regime, in order to make sure that the history-dependent QEW duration (Eq. 15) still satisfies the short wavepacket condition (Eq. 23) at the interaction point $z = 0$ after drift, and avoid slipping into the undesirable regime of anomalous FEBERI depicted in row C of Fig. 3. Then the natural dispersive broadening of the QEW (Eq. 15) sets a practical limit on the drift length $L_D$ before interaction. The minimum of the broadened QEW size $\sigma_t(L_D)$ with respect to $\sigma_{t0}$ is $\left(\sigma_t(L_D)\right)_{min} = \sqrt{L_D \lambda_c / 2\pi \gamma^3 \beta^3}$ (see supplementary Appendix D). Therefore, to satisfy the condition of a narrow Gaussian QEW $\Gamma = \omega_{2,1} \sigma_t(L_D) < \sqrt{2}$, one needs to limit the drift length before interaction to $L_D < z_G$ [36], where $z_G$ (Eq. 21) is a universal limit parameter of dispersive free-drift length, dependent only on the beam energy and the TLS transition frequency. In this paper, we use $\beta = 0.7, \omega_{2,1} = 3 \times 10^{15} Hz$ for all the presented results, for which $z_G = 4.9 cm$, an unprohibitive practical range for an electron microscope experimental setup.

## IV. INTERROGATION OF TLS WITH OPTICAL-FREQUENCY DENSITY MODULATED QEWS

In this section, we focus on the interaction between a TLS and an optical-frequency modulated QEW (both momentum and density modulations). In sub-section A we derive the development of density modulation in a finite size QEW after being momentum (energy) modulated in a PINEM process followed by free-space drift. In sub-section B we derive the FEBERI post-interaction spectrum of such a modulated QEW. We find good agreement between our numerical and analytical results and demonstrate the possibility of coherent control of the TLS-QS through density-modulated QEWs.

**A. Density modulation of a QEW**

Shaping of QEWs to sub-optical-cycle size can be done by a wavefunction chirping and drift process using THz or Infra-Red (IR) beams [16, 17]. An alternative way of shaping the QEW is by applying the PINEM process (Photon-Induced Near-Field Electron Modulation/Microscopy). In this process a broad QEW interacts with the near field of a laser beam illuminated structure, and gets energy modulated at the laser frequency $\omega_b$ by a process of multiple emission and absorption of photon quanta $\hbar \omega_b$ [19-23]. Such optical frequency modulation of QEWs has been demonstrated experimentally also by other laser-electron interaction schemes using dielectric structures, foils and the ponderomotive potential of laser beams beat [22, 25, 26, 44, 73]. It has been demonstrated experimentally that such an energy-modulated QEW becomes density modulated after drift at the laser modulation frequency and its harmonics, and produces an array of Attosecond scale current bunches [29, 33, 38]. Such short sub-bunches can play the same role as a near-point-particle single QEW [11]. Thus, in the case of FEBERI interaction with pre-modulated QEWs, the pre-shaping box in Fig. 1 represents a PINEM interaction setup, and $L_D$ is a drift distance at which the QEW develops density modulation of short (attosecond scale) sub-bunches.

The PINEM and the attosecond density modulation process have been analyzed in detail in several previous publications [19, 21, 74-76]. We iterate the exact derivation in Appendix D, and present here a simple useful approximation for the bunching coefficient in the practical case of a long QEW (relative to the optical modulation period). Though the physics of the FEBERI process is best comprehended intuitively in space-time coordinates, it is most straightforward to present the PINEM process in momentum space. Therefore, we start with the expression for the spectral sidebands structure of a PINEM-modulated Gaussian QEW in momentum presentation [19, 43]:

$$|\psi_p\rangle = \frac{1}{(2\pi\sigma_p^2)^{\frac{1}{4}}} \sum_m J_m(2|g_L|) \exp\left[-\frac{(p-m\delta p_L)^2}{4\sigma_p^2} - im\phi_0\right] |p\rangle \tag{24}$$

where $\delta p_L = \hbar\omega_b/v_0$ is the electron recoil momentum, $n$ is the order of an absorption or emission sideband, $\phi_0$ is the phase of the centroid of the QEW envelope relative to the near field generated by the laser at the PINEM modulation point [63] (see appendix F), $g_L$ is the coupling parameter of the interaction between the laser field and the electron ($g_L = \frac{e}{2\hbar\omega_L}\int E(z)e^{-i\Delta kz}\,dz$) [19], and it is assumed that the initial momentum (energy) spread of the QEW is small - $\sigma_p \ll \delta p_L$.

After drifting freely a distance $L_D$, the QEW in momentum representation is found by applying the evolution operator $e^{-iH_{0F}t_D/\hbar}$ with $H_{0F}$ being the second order expansion of the free space Hamiltonian (Eq. 4), and where $t_D = L_D/v_0$ is the free-drift traversal time of the electron:

$$c_p(t_D) = \frac{e^{-iE_p t_D}}{(2\pi\sigma_p^2)^{\frac{1}{4}}} \sum_m J_m(2|g_L|) e^{-\frac{(p-p_0-m\delta p_L)^2}{4\sigma_p^2} - im\phi_0} \tag{25}$$

where $E_p = E_0 + v_0(p-p_0) + \frac{1}{2\gamma^3 m_e}(p-p_0)^2$. Thus, we can derive the wavefunction in the spatial domain by Fourier transformation $\psi_z(z,t) = \int dp\, c_p(t)e^{ipz/\hbar}$, and the density probability distribution of the QEW is: $\rho_z = |\psi_z(z,t)|^2$. This space-time density modulation of the QEW was computed in Appendix F as a function of the drift length and the moving coordinate $z - v_0 t$ of the wavepacket envelope centroid, for the parameters presented in Fig. 4. The density distribution is composed of a train of attosecond scale short sub-bunches with periodicity (nearly) equal to the laser bunching period $T_b = 2\pi/\omega_b$, with a small frequency chirp due to the quadratic electron momentum (energy) dispersion. The modulation amplitude varies with $L_D$ in macroscopic scale (centimeters). It attains maximum modulation at $L_D = 1.5\ cm$ (Fig. 4b). The pattern repeats itself periodically (the temporal Talbot effect [77-79]).

An important figure of merit of the QEW density distribution in its interaction with radiation or other harmonic fields is the spectral bunching parameter:

$$\breve{\rho}_z(\omega) = \int_{-\infty}^{\infty} e^{i\omega t}\rho_z dt \tag{26}$$

This parameter, also termed the coherence coefficient [58, 80], is consistent with the corresponding definition of point-particle beam spectral bunching parameter in accelerator and FEL physics terminology [51, 81]. In the practical case of a long QEW envelope, the frequency chirp is small, and the density distribution may be written as a product of the envelope and density modulation functions [38] analogously to presentation in the classical bunched point-particle beam regime [82]:

$$\rho_z = f_{et}\left(t - \frac{z}{v_0} - t_0\right) f_{mod}\left(t - \frac{z}{v_0} - t_L\right) \tag{27}$$

where the envelope function is the absolute value square of the history-dependent broadened Gaussian QEW (Eq. 14) $f_{et}(t) = e^{-t^2/2\sigma_t^2(L_D)}/\left(2\pi\sigma_t^2(L_D)\right)^{1/2}$, $\omega_b t_L$ is the carrier envelope phase (CEP), and the modulation function is periodic and composed of harmonics of the bunching frequency:

$$f_{mod}\left(t - \frac{z}{v_0} - t_L\right) = \sum_{m=-\infty}^{\infty} b_m e^{m\omega_b\left(t - \frac{z}{v_0} - t_L\right)} \tag{28}$$

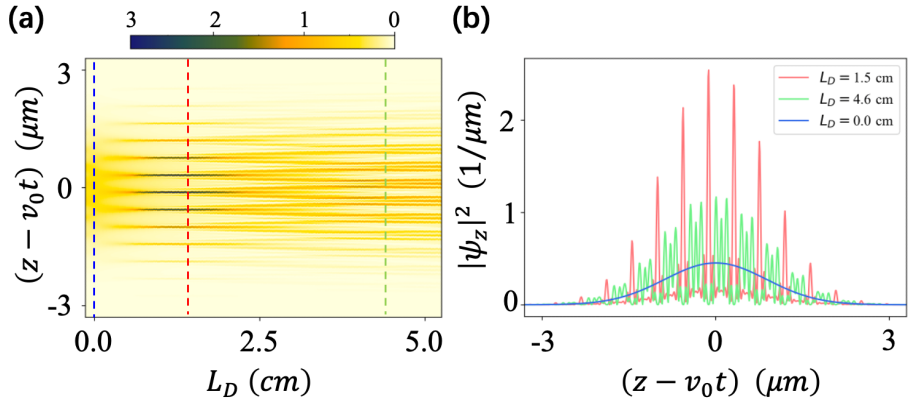

FIG. 4. Bunching of a modulated QEW. (a) Spatiotemporal evolution of the density distribution of a modulated QEW as a function of drift length; the blue vertical line at $L_D = 0\ cm$ marks the QEW energy modulation (PINEM interaction) point, the red line at $L_D = 1.5\ cm$ corresponds to maximum bunching (Eq. 31), and the green line at $L_D = 4.6\ cm$ corresponds to an over-bunching position. The beam energy is $E_0 = 205\ keV$, the (laser) bunching frequency is $\omega_b = 3 \times 10^{15}\ rad/s$, the envelope size of the QEW at $z = 0$ is $\sigma_{t0} = 2T_b = 4.2\ fs$, and the phase $\phi_0 = 0$. (b) The density distribution at different drift lengths. Continuous dependence of the density distribution on the drift length is shown in supplementary Movie S4 [70].

Thus, the spectral bunching parameter is

$$\breve{\rho}_z(\omega) = \sum_{m=-\infty}^{\infty} b_m e^{-\frac{(\omega - m\omega_b)^2 \sigma_t^2(L_D)}{2}} \tag{29}$$

where for the Gaussian modulated QEW (Eq. 25), the bunching parameter $b_m$ of harmonic frequency m can be expressed in absolute value by the simple expression (Appendix F) [58, 80]:

$$|b_m| = \left|J_m\left[4|g_L|\sin\left(\frac{2\pi m L_D}{z_T}\right)\right]\right| \tag{30}$$

This parameter attains its maximal value for the $m$-th harmonic $|b_m|_{max} = |J_m[u_m]|$, where $u_m = 4|g_L|\sin(2\pi m L_D^m/z_T)$ is the argument that makes the $m$-th order Bessel function maximal. The maximum bunching point is at a drift length location:

$$\left(\frac{L_D^m}{z_T}\right)_{max} = \frac{1}{2\pi}\sin^{-1}\left(\frac{u_m}{4|g_L|}\right) \tag{31}$$

where $z_T = 4\pi\beta^3\gamma^3 m_e c^3/\hbar\omega_b^2$ is the temporal Talbot distance [58, 80]. Interestingly enough, $z_T = 2\pi n^2 z_G$ for the resonant case $\omega_{2,1} = n\omega_b$, where $z_G$ was defined in Eq. (21) in connection to broadening of a finite Gaussian QEW in free space, both phenomena originate from the quadratic term in the energy dispersion relation ($E_p = E_0 + v_0(p - p_0) + \frac{1}{2\gamma^3 m_e}(p - p_0)^2$), analogous to the Fresnel diffraction effect in optics. Also interesting to observe that in the strong coupling limit $|g_L| \gg 1$, Eq. (31) reduces to $\left(\frac{L_D}{z_T}\right)_{max} = \frac{1}{4\pi|g_L|}$ for $m = 1$, which is the classical limit expression of maximal bunching [58, 80], and consistent with corresponding expressions for maximum point-particle beam bunching in accelerator and FEL theory [51, 81].

**B. FEBERI with a modulated QEW**

The analysis of the FEBERI interaction with a modulated QEW (Eq. 25) is similar to the analysis for a single finite size QEW, except that instead of (Eq. 13), one should substitute the modulated wavefunction expression (Eq. 25) in the generic expressions for the post-interaction free electron energy density distribution (Eq. 10). The initial momentum distribution in this case is $\rho_p^i = |c_p(t_D)|^2 = \frac{1}{\sqrt{2\pi\sigma_p^2}}\sum_m |J_m(2|g_L|)|^2 e^{-\frac{(p-p_0-m\delta p_L)^2}{2\sigma_p^2}}$. The zero, first and second order terms in the post-interaction incremental density distribution (Eq. 16) are derived in Appendix G under the assumption of near resonance condition of a harmonic order $n$ of the modulated QEW with the TLS transition $n\omega_b \approx \omega_{2,1}$:

$$\Delta\rho_p^{(0)} = -\sin^2|g|\sum_m |J_m(2|g_L|)|^2 \frac{1}{\sqrt{2\pi\sigma_p^2}} \exp\left[-\frac{(p-p_0-m\delta p_L)^2}{2\sigma_p^2}\right] \tag{32}$$

$$\Delta\rho_p^{(1)} = \frac{\sin(2|g|)}{2}\sin\theta\, e^{-\frac{(\omega_{2,1}-n\omega_b)^2\sigma_{t_0}^2}{2}}\sum_m A_{n,m}(\phi)\frac{1}{\sqrt{2\pi\sigma_p^2}}\exp\left[-\frac{(p-p_0-m\delta p_L)^2}{2\sigma_p^2}\right] \tag{33}$$

$$\Delta\rho_p^{(2)} = \sin^2|g|\sum_m B_{n,m}(\theta)\frac{1}{\sqrt{2\pi\sigma_p^2}}\exp\left[-\frac{(p-p_0-m\delta p_L)^2}{2\sigma_p^2}\right] \tag{34}$$

where $A_{n,m}(\phi) = \left[-\sin(\zeta + \phi_g + \phi_{\delta p_L}^m)J_{m+n}(2|g_L|) + \sin(\zeta + \phi_g + \phi_{-\delta p_L}^m)J_{m-n}(2|g_L|)\right]J_m(2|g_L|)$ is the first order incremental amplitude of the sideband order $m$ in $n$-th order resonant interaction with the TLS (Eq. 2), $B_{n,m}(\theta) = |J_{m-n}(2|g_L|)|^2 \cos^2\left(\frac{\theta}{2}\right) + |J_{m+n}(2|g_L|)|^2 \sin^2\left(\frac{\theta}{2}\right)$ is the second order incremental amplitude of the sideband order $m$, $\zeta = \omega_{2,1}t_D - \phi + n\phi_0$ is the relative phase between the modulated QEW and the TLS excitation phase. The additional phase $\phi_{\pm\delta p_L}^m = \omega_{2,1}t_D\frac{(m\pm n/2)\delta p_L}{2\gamma^3 m v_0}$ is the consequence of the difference in electron recoil due to positive or negative transfer of m quanta in the quantum transition of the TLS.

Equations (32-34) and Figs. (5a) and (5b) show that after interaction, the multiple sidebands structure of the PINEM-modulated QEW remains peaked at $p = p_0 + m\delta p_L$, but the momentum density distribution is redistributed due to FEBERI induced recoils between the sideband orders. Similar to the case of unmodulated QEW, the zero-order modification of the density distribution is negative and independent of the TLS-QS qubit parameters, the second order term depends on $\theta$ only, and the first order term depends on both $\theta$ and $\phi$ (through $\zeta$). Evidently, with $|g \ll 1|$, the zero-order and second-order terms (proportional to $|g|^2$) are much smaller than the first-order term (proportional to $|g|$), so $\Delta\rho^{(1)}$ is the dominant term for the incremental spectrum, except near the poles of the Bloch sphere.

We first consider the case without drift, $L_D = 0$ or short drift, such that the recoil phase shift is $\phi^m_{\pm\delta p_L} \ll 1$. In this case we set $\phi^m_{\pm\delta p} = 0$ and

$$A_{n,m} = [-J_{n+m}(2|g_L|) + J_{m-n}(2|g_L|)]J_m(2|g_L|) \sin(\zeta + \phi_g) = (-1)^{n+1} A_{n,-m}$$

Therefore, the sidebands of the incremental distribution $\Delta\rho_p^{(1)}$ are distributed symmetrically around $p = p_0$ for n=odd, and antisymmetrically for n=even. The dependence on $\theta$ and $\phi$ (through $\zeta$) is simply sinusoidal, in proportion to $\sin\theta \sin(\zeta + \phi_g)$, as shown for the case n=1 in Figs. (5a) and (5b) respectively, with good agreement between the analytical and numerical computation.

The case of finite drift $L_D \neq 0$, and specifically $L_D = (L_D)_{max} = 3.3\ cm$ is displayed in Figs. (5c-5f). Figures (5c) and (5d) depict the dependence of the post-interaction QEW energy density spectrum as a function of the azimuthal $\phi$ and polar $\theta$ qubit parameters. The sidebands are redistributed asymmetrically in this case. Consequently, the electron energy increment – the integral of the energy density distribution increment – presents net acceleration dependence on $\theta$ and acceleration/deceleration dependence on $\phi$, as depicted in Figs. (5e) and (5f) respectively, showing full agreement between the analytical and numerical results.

Complementary to the explicit expressions for the incremental free electron energy spectrum (Eqs. 32-34), the comprehensive model solution of Schrodinger equation for the entangled free and bound electrons also provides the corresponding expression for the incremental TLS transition probability (see Appendix G):

$$\begin{aligned}\Delta P_2 = -\Delta P_1 = &- \sin^2|g| \cos\theta - \frac{1}{2}\sin(2|g|) \sin\theta \sin\left(\zeta + \phi_g + 2\pi n \frac{L_D}{z_T}\right) \\ &\times J_n\left(4|g_L| \sin\left(2\pi n \frac{L_D}{z_T}\right)\right) exp\left[-\frac{(\omega_{2,1} - n\omega_b)^2 \sigma_{t0}^2}{2}\right] e^{-\Gamma_D^2/2}\end{aligned} \quad (35)$$

The energy change of the TLS - $E_{2,1} \cdot \Delta P_2$, is depicted in Figs. (5e) and (5f) together with the change of the free electron energy, showing full agreement with the conservation of energy relation (Eq. 12), thus affirming the self-consistence of the formulation and the validity of using the post-interaction energy spectrum of modulated QEWs for diagnosis of the TLS-QW qubit parameters.

Equation (35) is fully consistent with Eq. (30) for the bunching parameter of a density modulated QEW, showing that the first order energy transfer between the electron and the TLS is proportional to the bunching coefficient $b_n$. This fundamental relation between free electron bunching and energy transfer is analogous to similar relations in classical electron interaction with light (such as in accelerators and FEL [51, 82]).

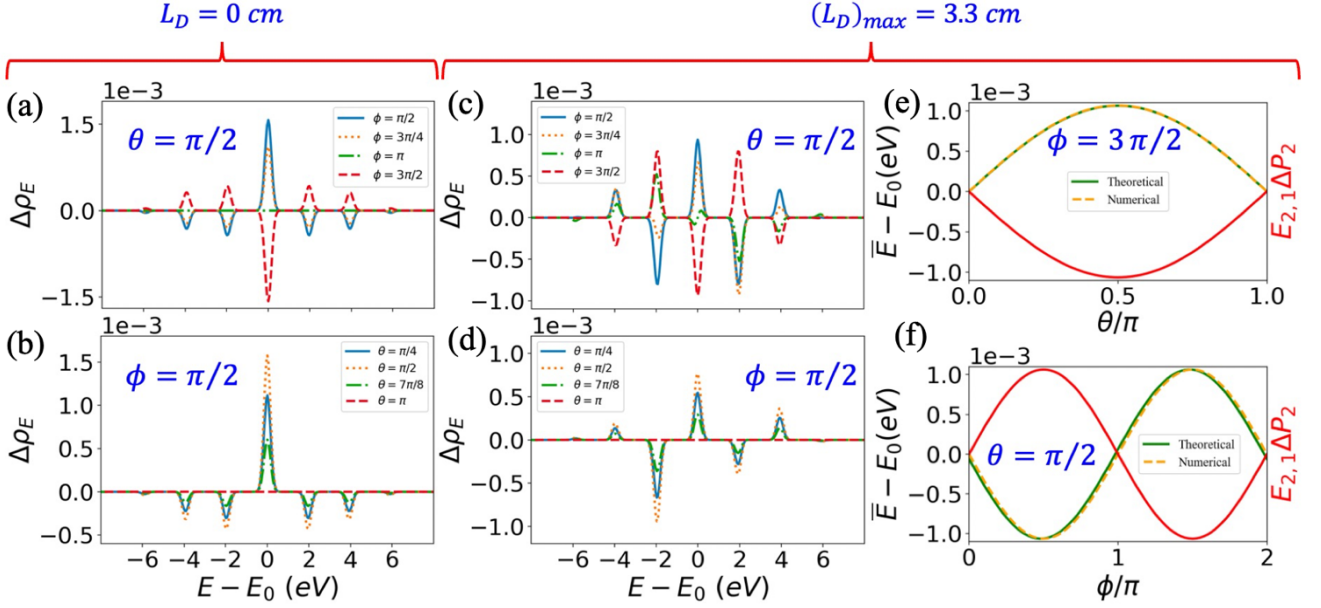

FIG. 5. Incremental energy spectra and acceleration/deceleration of a modulated QEW. The incremental post-interaction spectrum of an energy modulated QEW without bunching ($L_D = 0$) depends on the azimuth angle $\phi$ (a) and polar angle $\theta$ (b). The corresponding dependences on $\phi$ and $\theta$ of the incremental spectral density distributions of a density bunched modulated QEW ($L_D \neq 0$) are shown in (c) and (d), respectively. The acceleration/deceleration curves of the bunched QEW vary with polar angle $\theta$ (e) and azimuth angle $\phi$ (f). The corresponding energy change $E_{2,1} \cdot \Delta P_2$ of the TLS is also shown in (e) (f) and satisfies the energy conservation relation (Eq. 12). There is a full agreement between the analytical and numerical results, which were computed for wavepacket size $\sigma_{t0} = T_{2,1}$, phase $\phi_0 = 0$, and PINEM modulation intensity $2|g_L| = 1.5$.

It is instructive to note that substitution of $L_D = 0$ (or $|g_L| = 0$) diminishes all the harmonics except $n = 0$ and reduces Eq. (35) to the expression for an unmodulated Gaussian QEW (Eq. 20). When $L_D = 0$, there is significant first order incremental change in all sidebands of the energy spectrum (Eq. 33, Figs. 5a, 5b), however, there are no first order incremental transitions of the TLS (the second term in Eq. 35 vanishes). Finite first order transition probability of the TLS and corresponding acceleratin/deceleration of the QEW are possle only when $L_D \neq 0$. Their dependence on $(\phi, \theta)$ is depicted in Figs. (5e) and (5f). The physical explanation for that is that only then density modulation takes place, and in time-space domain the QEW sub-bunches acquire phase, and can interact with the TLS like near-point-particles in correspondence to the Born interpretation of the probability density of the QEW. Note that in the case of a modulated QEW (Figs. 5a-5d), the spectral peaks of the energy spectrum are three orders of magnitude larger than with a long unmodulated QEW (Fig. 2d). This is true for both cases of $L_D = 0$ and $L_D \neq 0$, However in the first case no first order quantum transitions take place, and the TLS-QS stays nearly intact, and in the second case there is first order change of the ocupation probabilities of the TLS and the polar angle of the qubit in the Bloch sphere. For this reason it is preferable to use the short drift case for non-destructive probing of the TLS-QS, and the maximum bunching finite drift length case - for coherernt quantum control.

## V. OFF RESONANCE FEBERI AND TEMPORAL EVOLUTION OF TLS TRANSITIONS

Except for the small second-order term, the dominant first order term in Eq. (35) indicates that the FEBERI process of a modulated QEW requires satisfaction of a resonance condition (Eq. 2) in order to attain controlled TLS quantum transitions. However, this resonance requirement is not sharp when the size of the envelope of the QEW $\sigma_{z0}$, is finite. This is also depicted by the numerically computed curves in Fig. 6 that are in full agreement with the analytical expression. The shorter the size of the QEW envelope, the wider is the tolerance of transition probability to off-resonance interaction, however the envelope size must be kept short enough to keep $\Gamma_D < 1$, to avoid the exponential decay in Eq. (35).

It is intriguing to get deeper insight into the off-resonance decay of the transition probability by examining the temporal evolution of the transition probability during the interaction period of the modulated QEW with the TLS. The numerically computed incremental occupation probability of the upper TLS level is shown in Fig. 7 as a function of interaction time comparatively for the case of resonant and off-resonant interactions. Comparing the curves of temporal dependence of TLS quantum transition probability (Fig. 7b) to the density probability curves of the modulated QEW (Fig. 7a), we see that the "quantum jumps" of the TLS take place at times commensurate with the period of the quantum transition frequency $T_{2,1} = 2\pi/\omega_{2,1}$ whether the modulated QEW is at resonance or not. When the bunching is out of resonance, the increments of quantum jump probability diminish towards the end of the interaction time, and consequently the measurable post-interaction incremental transition probability is smaller than in resonance setting. Evidently, this happens because the QEW bunching gets out of phase with the natural dipole oscillation of the TLS.

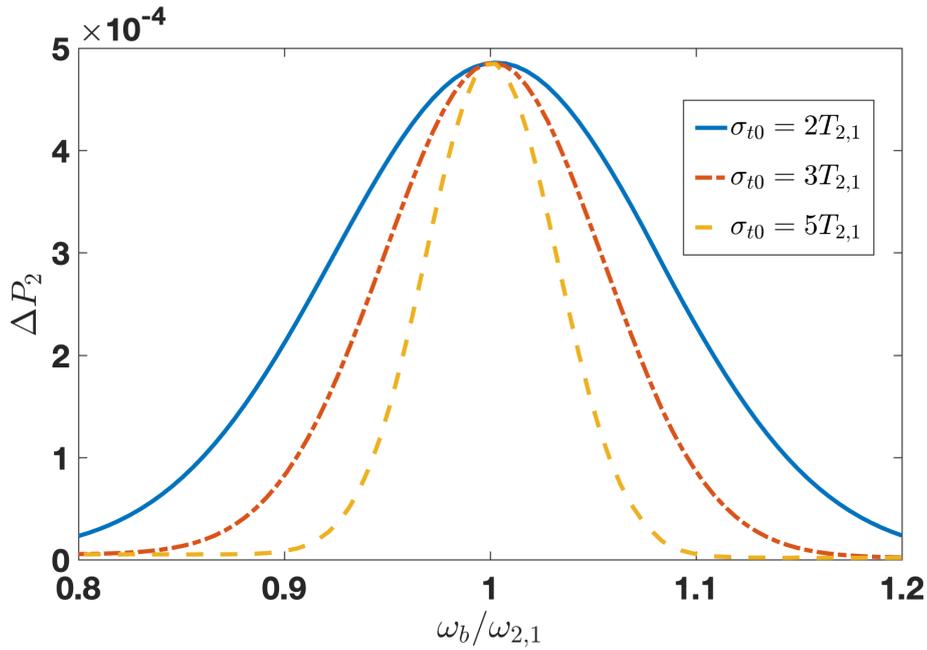

FIG. 6. Resonance of the FEBERI-induced quantum TLS transition probability. The post-interaction incremental transition probability $\Delta P_2$ dependence on the ratio of the bunching (laser) frequency and TLS transition frequency $\omega_b/\omega_{2,1}$ is depicted for different wavepacket envelope sizes. The curves were computed for pre-interaction PINEM modulation intensity $2|g_L| = 1.5$.

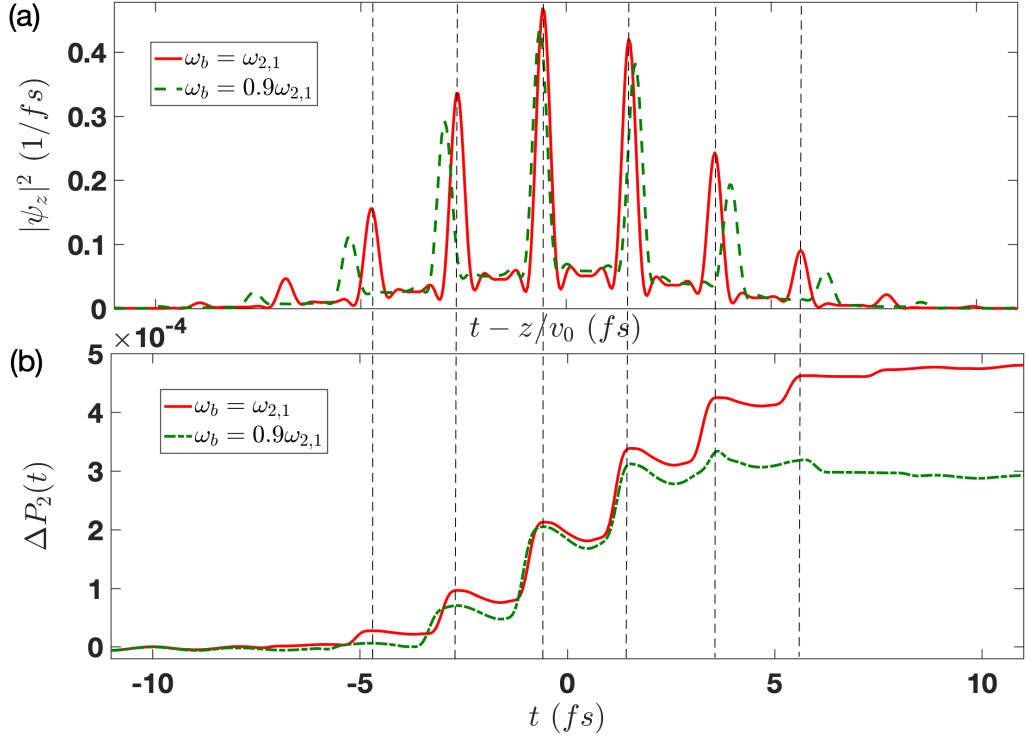

FIG. 7. Evolution of transition probability induced by a density modulated QEW. (a) depicts the spatio-temporal density distribution of the QEW, and (b) depicts the corresponding incremental transition probability of the TLS, both for resonant and off-resonant modulation cases. The time dependent upper state incremental probability is $\Delta P_2(t) = P_2(t) - P_2^i$, where $P_2(t)$ is the numerical solution of Eq. (H5), derived from the combined free-bound electrons Schrodinger equation (Eq. 6). The curves were computed for modulation intensity $2|g_L| = 1.5$ and wavepacket size $\sigma_{t0} = 1.5T_{2,1}$.

It is instructive to examine the quantum transition process from the point of view of Born's interpretation of the electron wavefunction [72]. This is possible in the present theory since the quantum wavepacket formulation encompasses both limits of the particle-wave duality in spatiotemporal presentation. First consider a finite size (unmodulated) QEW (Eq. 20). In the wavelike limit of a long QEW and small energy spread (Eq. 22), the first order transition probability decays exponentially, and what is left in the transition probability (Eq. 20) is only the second order term, which is independent of the azimuth angle $\phi$ of the qubit. On the other hand, in the near-point-particle limit of the QEW (Eq. 23), the dominant term in the transition probability (Eq. 22) is the azimuth-phase-dependent first order term. This is Consistent with the probabilistic Born interpretation, since in the latter case there is high probability for arrival of the QEW at the FEBERI interaction point with well-defined phase relative to the oscillating dipole moment of the TLS.

The case of a density-modulated QEW is more ambiguous. On one hand, the envelope of the modulated QEW is long relative to the modulation period, but on the other hand, the duration of the density modulation sub-bunches (attosecond scale – see Fig. 4) satisfies the short wavepacket condition (Eq. 23). Since we deal here with a single electron, the sub-bunches do not represent individual electrons as in the case of the semiclassical "optical klystron" [54]. Instead, all the sub-bunches in the time-space distribution shown in Fig. (4b) represent a single electron. According to the Born interpretation, the electron has highest probability to arrive to the FEBERI interaction point at a time corresponding to one of the peaks of the sub-bunches and appear as a near-point-particle QEW. The analytical expression (Eq. 35) and Fig. 7 affirm the intuitive interpretation of the sub-bunches as near-point-particle wavepackets. Evidently, the "quantum jumps" temporal evolution curves of Fig. 7 do not represent the evolution of the TLS due to passage of a single electron, but only the expectation value of $P_2$ over multiple modulation-phase correlated QEWs, whereas in each individual FEBERI event, only one of the sub-bunches of the QEW is responsible for the "quantum jump". With multiple correlated electrons, these increment jumps of the transition probability would take place in synchronization with the TLS dipole oscillation frequency $\omega_{2,1}$. This happens around the random times, of the QEWs centroid arrival within the QEW envelope duration. This observation is crucial for the generalization of the FEBERI effect to the case of multiple modulation-correlated QEWs, where coherent build-up of TLS transitions proportional to $N^2$ is expected [11, 48, 55, 56]. It gives a lead to a Born interpretation inspired concept, in which the joint wavefunction of a pulse of modulation-correlated QEWs (e.g., as proposed in [57]) can interrogate and coherently control the TLS-QS collectively.

## VI. DISCUSSIONS AND CONCLUSIONS

In this article we present a scheme for diagnostics and coherent control of the superposition quantum state of a TLS in a variety of possible targets of interest, including, but not limited to atoms, quantum dots and crystal electron defect centers. Such targets can be addressed by an electron beam with atomic-scale spatial resolution. Hence, this scheme may be useful for such applications as full characterization of the TLS state as a multi-qubit on a single Bloch sphere in quantum information technology [6, 7], measurement of decoherence time [61] and selective coherent excitation of quantum emitters [66].

The interrogation scheme is based on the recently proposed FEBERI process [11]. The electron quantum wavefunction is pre-shaped after its emission from the cathode by coherent processes of optical chirping and streaking at THz or IR frequencies, followed by drift [16, 17]. This causes it to arrive at the TLS location in a controlled shape – longer or shorter than the TLS transition cycle $T_{2,1} = 2\pi/\omega_{2,1}$ (of the order of femtosecond). Alternatively, the QEW may be energy modulated before the FEBERI by a laser beam in a PINEM interaction process at a sub-harmonic frequency of the TLS transition frequency $\omega_{2,1}$. After free drift, the energy modulated QEW becomes periodically density-modulated (bunched) at the PINEM modulation frequency with tight (attosecond) sub-bunches, and contains high frequency spatio-temporal harmonics of the modulation frequency. In either case, the FEBERI interaction with an excited TLS-QS results in a post-interaction spectrum of multiple repeated measurements that reflects the quantum state of the TLS. This can be used for diagnostics of the TLS-QS qubit parameters on the Bloch sphere, and for coherent control of the TLS-QS.

Determination of the polar angle $\theta$ of the TLS-QS can be carried out by using narrow energy spread QEWs that satisfy the large recoil condition $E_{2,1} > 2\sigma_E$ (Eq. 22, Fig. 3a). This requires sub-eV electron energy spread, which is readily available in electron microscopy. The post-interaction incremental energy density distribution of such an electron beam depicts a PINEM-kind spectrum of two asymmetric sidebands (Fig. 2d, Movie S2). The degree of asymmetry indicates the polar angle $\theta$ of the pre-set interrogated TLS-QS. The determination of the TLS-QS azimuthal angle $\phi$ is less straightforward. It necessitates forming narrow QEWs that satisfy the near-point-particle limit condition $2\sigma_t(L_D) < T_{2,1}$ (Eq. 23, Fig. 3b) at the FEBERI interaction location. The resulting post-interaction incremental energy density distribution depicts in this limit an acceleration/deceleration spectrum that depends on the phase increment of the TLS-QS, accumulated from the time of TLS pre-excitation up to the arrival time of the QEW centroid: $\zeta = \omega_{2,1} t_D - \phi - \varphi_L$. For this measurement, the TLS excitation and the timing of the QEW beam must be phase locked, and the TLS-QS azimuth angle $\phi$ can be then determined by judicial adjustment of the optical path difference between the two laser beams that determines the reference phase $\varphi_L$ (see Fig. 1).

The alternative proposed method for interrogating the TLS-QS using optical frequency modulated QEWs is expected to be more accessible in the lab. It requires energy modulation of the QEWs before the FEBERI point by means of the well-established PINEM process. We have demonstrated that such modulated QEWs may be used for interrogating the TLS-QS qubit parameters $(\theta, \phi)$ in two different modes: without drift after PINEM ($L_D = 0$) and with finite drift ($L_D \neq 0$). In the first case the QEW is only energy modulated (PINEM sidebands) and there is no density modulation, the post-interaction incremental spectrum is either symmetric (for $n = odd$) or antisymmetric $n = even$, and is dependent on both $(\theta, \phi)$ (see Figs. 5a, 5b). There is no net energy transfer between the TLS and the QEW in this case. In the second case, density bunching of the modulated QEW is formed due to the pre-interaction free drift, and the post-interaction incremental spectrum is asymmetric and dependent on $(\theta, \phi)$ (Figs. 5c, 5d). There is then net energy transfer between the QEW (that is accelerated or decelerated – see Figs. 5e, 5f) and the TLS. Consequently, the TLS changes its quantum levels occupations ratio (measured by the qubit parameter $\theta$).

In both cases of modulated QEW, with and without drift, the incremental spectrum signal is significantly larger than in the case of an unmodulated long QEW (for the parameters used – three orders of magnitude more in comparing Figs. (5a-5d) to Figs. (2d), (2e). It is however of the same order of magnitude as in the case of a short QEW (Fig. 2a). The reason is that the azimuthal phase-dependent first order term ($\Delta\rho^{(1)} \propto |g|$) dominates in both cases over the second order term ($\Delta\rho^{(1)} \propto |g|^2$) for $|g| \ll 1$. Further, we draw attention to the fact that there is energy transfer only in the case of short sub-cycle QEW (Fig. 2b, 2c) and the case of modulated QEW after drift $L_D \neq 0$ (Figs. 5e, 5f). This should be related to the development of bunching with attosecond-scale sub-bunches after drift, as shown in Fig. 4. In the spatio-temporal presentation, we are led by Born's interpretation of the quantum electron wavefunction to conclude that these sub-bunches can be considered as multiple near-point-particles at the FEBERI interaction point. Therefore, similarly to sub-cycle QEWs in the near-point-particle limit, they have well defined phase relative to the phase $\phi$ of the TLS dipole oscillation, and in both cases the phase dependent first order term ($\Delta\rho^{(1)} \propto |g|$) in (Eq. 20, Eq. 35) facilitates the transfer of energy between the QEW and the TLS. This observation and the analysis of the temporal development of TLS transitions in Fig. 7 of the previous section, have important ramification to the extension of the single modulated QEW FEBERI concept to multiple modulation-correlated QEWs [11, 48].

The laboratory realization of the proposed concept of TLS-QS interrogation and coherent control using free electrons, faces substantial theoretical and conceptual challenges as well as technological implementation challenges that we address below. In the first place, any projective measurement of a TLS quantum state by EELS technique is destructive upon every EELS data measurement event. Therefore, as we indicated in section II, the measurement of our predicted post-interaction energy density distribution by EELS requires reinstating the TLS-QS after each EELS measurement data collection event in a way similar to experimental practice in super-elastic scattering research [83]. This limits the application of the proposed scheme to cases where the TLS-QS targets are reconstructable, and the multiple QEWs are identical. For example, we conceive an application of monitoring the Bloch sphere time-evolution trajectory of laser beam coherent control of a TLS [84] with an electron beam. Realization of such an application requires that the laser coherent control process is reinstated identically after every destructive EELS data collection by the interrogating QEWs. We exclude consideration of nondestructive measurements. These have been demonstrated and proposed with photons [85, 86] and electrons [87].

On top of this fundamental limitation on EELS measurement, there are also technical impediments on realizing a true energy spectral distribution by means of multiple single electron EELS measurements. In the first place, the cross section for an inelastic scattering event is very small (in our examples, $\sim 10^{-6}$ per electron for a broad unmodulated QEW, and $\sim 10^{-3}$ per electron for a modulated QEW). Another limitation is that any single event measurement of the qubit parameters $(\phi, \theta)$ must take place within the decoherence time $T_2$ and the relaxation time $T_1$. One can conceive a scheme of accumulating EELS data with a pulse of multiple QEWs interacting with multiple TLS centers, excited all with a laser pulse to the same state (for example, employing multiple crystal defect centers implanted on a crystal surface). However, the single atom resolution advantage of the FEBERI scheme is then lost, and for the same reason, there is no possibility then of interrogating the azimuthal phase $\phi$, only the average polar angle of the TLS ensemble can be interrogated.

Further investigations are needed for identifying TLS targets of bigger dipole moment and coupling coefficient $g$, e.g., perovskite [88]. We point out here also to another potential scheme for enhancing the FEBERI interaction process and amplifying the spectral energy distribution signal in the interrogation of a TLS-QS. In this scheme one would place multiple ($N_B$) TLS targets within a volume smaller than $v_0/\omega_{2,1}$, such that they can all be excited in phase to the same quantum state by the exciting laser. In this case, a pulse of uncorrelated QEWs interacting each with a single TLS in the lump of TLS can produce multiple EELS data points without destruction of the qubit state of other TLS targets in the lump. Furthermore, if the TLSs are lumped within the range of the QEWs field, they may act as a "super-qubit" with dipole moment and interaction coupling parameter $g$ that are $N_B$ times bigger, and then the EELS probing signal, as well as the transition probability, would be enhanced correspondingly. This situation is similar to the concept of superradiant emission, induced by free electrons [64]. In fact, besides enhancing the EELS measurement, such an excited multiple-TLS "super-qubit" is expected to emit superradiant luminescence proportional to $N_B^2$ after the radiative relaxation lifetime of the TLS [61].

With a "super-qubit" of many TLS targets, it is possible to keep the high spatial resolution addressing advantage of a focused electron beam. However, if the electrons of the beam are uncorrelated, it is still necessary to reinstate the TLS-QS after each measurement event, in order to collect enough data for constructing the EELS. The "Holly Grail" of TLS-QS interrogation by FEBERI would be achieved if a pulse of optical frequency modulation-correlated QEWs can be produced within the duration of the decoherence or relaxation time. In this case one may conceive that the joint wavefunction of a pulse of modulation-correlated QEWs (e.g., as proposed in [48]) can interrogate and coherently control the TLS-QS collectively, similarly to a laser beam, but with high spatial resolution. This concept requires further intended theoretical considerations beyond the scope of the present publication (for example of the entanglement between the interacting electrons), and certainly further technological development.

Finally, we draw attention to the rapid development of the field of quantum electron-optics and the marriage of integrated photonics with electron microscope technology, such as enhanced coupling of light through micro-resonators [89, 90, 91] and embedding single ions in a nanophotonic cavity [7]. These have the potential of bringing about new development directions of the FEBERI concept, providing more degrees of freedom in control and measurement of electron-light-TLS interaction. For example, we may envision schemes for extension of TLS decoherence time by suppression of spontaneous emission in a microcavity and electron-light interaction schemes with multi-cavity, multi-TLS integrated photonics circuits.

## Acknowledgments

We acknowledge helpful communications with O. Kfir, Y. M. Pan and F. J. Garcia de Abajo. This work was supported by the Israel Science Foundation (ISF) under Grant No. 00010001000, the National Natural Science Foundation of China under Grant No. 12104068, the Natural Science Foundation of Chongqing under Grant No. cstc2021jcyj-msxmX0684. D.R. acknowledges support by the PBC program of the Israel Council of Higher Education.

**Author contributions:** A.G., A.Y., J.S., and D.R. conceived the concept. B.Z. and D.R. performed the theoretical derivations and prepared the figures and videos. B.Z., D.R. and A.G. wrote the paper with contributions from R.I. and A.F. All authors reviewed and discussed the manuscript and made significant contributions to it.

## APPENDIX A: DERIVATION OF THE INTERACTION HAMILTONIAN $\widehat{\mathcal{H}}_I(t)$ IN THE INTERACTION PICTURE

In this section, we elaborate the derivation of the interaction Hamiltonian in the interaction picture. For the model of free electron–bound electron interaction, the dominant interaction is the Coulomb interaction. Within the dipole approximation, the interaction Hamiltonian is $\widehat{H}_I = e\hat{\boldsymbol{r}}' \cdot \boldsymbol{E}(\hat{z}, r_{\perp 0})$, where $\hat{\boldsymbol{r}}'$ is the position vector of the bound electron, and $\boldsymbol{E}(\hat{z}, r_{\perp 0}) = \frac{e\gamma}{4\pi\varepsilon_0} \frac{z\hat{e}_z + r_{\perp 0}\hat{e}_\perp}{(\gamma^2 z^2 + r_{\perp 0}^2)^{3/2}}$ is the electric field generated by the free electron at the position $\hat{\boldsymbol{r}}'$.

The interaction Hamiltonian in the interaction picture is defined as:

$$\begin{aligned}\widehat{\mathcal{H}}_I(t) &= \widehat{U}^\dagger(t)\widehat{H}_I\widehat{U}(t) \\ &= e^{i\widehat{H}_{0B}t/\hbar}(e\hat{\boldsymbol{r}}')e^{-i\widehat{H}_{0B}t/\hbar} \times e^{i\widehat{H}_{0F}t/\hbar}\boldsymbol{E}(z, r_{\perp 0})e^{-i\widehat{H}_{0F}t/\hbar} \\ &= \left[\sum_{ij}|i\rangle\langle i|e^{\frac{i\widehat{H}_{0B}t}{\hbar}}(e\hat{\boldsymbol{r}}')e^{-\frac{i\widehat{H}_{0B}t}{\hbar}}|j\rangle\langle j|\right]e^{i\widehat{H}_{0F}t/\hbar}\boldsymbol{E}(z, r_{\perp 0})e^{-i\widehat{H}_{0F}t/\hbar}\end{aligned} \quad (A1)$$

where $\widehat{H}_{0B}$ is the Hamiltonian of the bound electron and $|i\rangle, |j\rangle$ are the eigenstates of $\widehat{H}_{0B}$. $\widehat{H}_{0F} = E_0 + v_0 \cdot (\hat{p} - p_0) + \frac{1}{2\gamma^3 m}(\hat{p} - p_0)^2$ is the kinetic Hamiltonian of a free electron, which is valid in the relativistic case. Eq. (A1) can be rewritten in a matrix form:

$$\widehat{\mathcal{H}}_I(t) = \begin{pmatrix} 0 & \boldsymbol{\mu}_{2,1}^* e^{i\omega_{2,1}t} \\ \boldsymbol{\mu}_{2,1} e^{-i\omega_{2,1}t} & 0 \end{pmatrix} e^{i\widehat{H}_{0F}t/\hbar}\boldsymbol{E}(z, r_{\perp 0})e^{-i\widehat{H}_{0F}t/\hbar} \quad (A2)$$

where $\boldsymbol{\mu}_{2,1} = e\langle 2|\hat{\boldsymbol{r}}'|1\rangle$ is the dipole moment of the TLS and $\omega_{2,1} = \omega_2 - \omega_1$ is the energy difference between the two levels. Since the interaction time is short, the quadratic term in $\widehat{H}_{0F}$ is negligible, leading to

$$\begin{aligned}\widehat{\mathcal{H}}_I(t) &= \begin{pmatrix} 0 & \boldsymbol{\mu}_{2,1}^* e^{i\omega_{2,1}t} \\ \boldsymbol{\mu}_{2,1} e^{-i\omega_{2,1}t} & 0 \end{pmatrix} e^{iv_0\hat{p}t/\hbar}\boldsymbol{E}(z, r_{\perp 0})e^{-iv_0\hat{p}t/\hbar} \\ &= \begin{pmatrix} 0 & \boldsymbol{\mu}_{2,1}^* e^{i\omega_{2,1}t} \\ \boldsymbol{\mu}_{2,1} e^{-i\omega_{2,1}t} & 0 \end{pmatrix} \boldsymbol{E}(z + v_0 t, r_{\perp 0}) \\ &= \boldsymbol{\mu}_{2,1} \cdot \boldsymbol{E}(z + v_0 t, r_{\perp 0})e^{-i\omega_{2,1}t}\sigma^- + h.c.\end{aligned} \quad (A3)$$

where the operator $e^{iv_0\hat{p}t/\hbar}$ is the translation operator in coordinate domain, and $\hat{\sigma}^+ = |2\rangle\langle 1|$ ($\hat{\sigma}^- = |1\rangle\langle 2|$) is the TLS raising (lowering) operator.

# APPENDIX B: DERIVATION OF THE SCATTERING OPERATOR $\hat{S}(t_f, t_i)$

With the explicit form of the interaction Hamiltonian, we can calculate the wavefuction after the interaction by multiplying the initial wavefunction with the scattering operator:

$$|\widetilde{\Psi}_f(p, t_f)\rangle = \hat{S}(t_f, t_i)|\widetilde{\Psi}_i(p, t_i)\rangle$$
$$\hat{S}(t_f, t_i) = \mathcal{T} \exp\left[-\frac{i}{\hbar}\int_{t_i}^{t_f} \widehat{\mathcal{H}}_I(t)dt\right] \tag{B1}$$

where $\mathcal{T}$ is the time-ordering operator. The scattering operator can be expanded to a Dyson series

$$\hat{S}(t_f, t_i) = \sum_{n=0}^{+\infty} \hat{S}_n(t_f, t_i) \tag{B2}$$

where

$$\hat{S}_n(t_f, t_i) = \exp\left[\frac{(-i/\hbar)^n}{n!}\int_{t_i}^{t_f} dt_1 \int_{t_i}^{t_f} dt_2 \ldots \int_{t_i}^{t_f} dt_n \, \mathcal{T}\widehat{\mathcal{H}}_I(t_1)\widehat{\mathcal{H}}_I(t_2)\ldots\widehat{\mathcal{H}}_I(t_n)\right] \tag{B3}$$

The first order term in this series is easily obtained:

$$\hat{S}_1(t_f, t_i) = \exp\left[-\frac{i}{\hbar}\int_{t_i}^{t_f} \widehat{\mathcal{H}}_I(t)dt\right]$$
$$= \exp\left[-\frac{i}{\hbar}\int_{-\infty}^{+\infty} dt[\boldsymbol{\mu}_{2,1} \cdot \boldsymbol{E}(z + v_0 t, r_{\perp 0})e^{-i\omega_{2,1}t}\sigma^- + h.c.]\right] \tag{B4}$$

Replacing the integral variable t by $u = z + v_0 t$ yields

$$\hat{S}_1(t_f, t_i) = \exp[-\frac{i}{\hbar v_0}\int_{-\infty}^{+\infty} du[\boldsymbol{\mu}_{2,1} \cdot \boldsymbol{E}(u, r_{\perp 0})e^{-i\omega_{2,1}u/v_0}\sigma^- e^{i\omega_{2,1}z/v_0} + h.c.]] \tag{B5}$$

The integration on the field presents the coupling coefficient between the free electron and the TLS: $g = \frac{1}{\hbar v_0}\boldsymbol{\mu}_{2,1} \cdot \int_{-\infty}^{\infty} du \boldsymbol{E}(u, r_{\perp 0})e^{-i\omega_{2,1}u/v_0}$. Substituting the field $\boldsymbol{E}(\hat{z}, r_{\perp 0})$ into (Eq. S8) yields:

$$g = \frac{e}{2\pi\epsilon_0\gamma\hbar}\frac{\omega_{2,1}}{v_0^2}\boldsymbol{\mu}_{2,1} \cdot \left[\frac{i}{\gamma}K_0\left(\frac{\omega_{2,1}r_{\perp 0}}{v_0\gamma}\right)\hat{\boldsymbol{e}}_z + K_1\left(\frac{\omega_{2,1}r_{\perp 0}}{v_0\gamma}\right)\hat{\boldsymbol{e}}_\perp\right] = |g|e^{i\phi_g} \tag{B6}$$

Considering laboratory available parameter values - TLS energy gap of $E_{2,1} = \hbar\omega_{2,1} = 1.97\ eV$, free electron velocity of $v_0 = 2.1 \times 10^8\ m/s$ (β = 0.7) and impact parameter $r_{\perp 0}$= 2.0 nm, we find the coupling coefficient g to be in the order of $10^{-3}$. Thus, it is reasonable to neglect the higher order terms in Dyson series. Hence, the scattering matrix can be written as

$$\hat{S}(t_f, t_i) = cos|g| - i\, sin|g|\, \left(\hat{\sigma}^+ e^{-i\omega_{2,1}\hat{z}/v_0} e^{i\phi_g} + h.c.\right) \tag{B7}$$

## APPENDIX C: DENSITY DISTRIBUTION AND TRANSITION PROBABILITY AFTER INTERACTION

For any input state $|\tilde{\psi}_i\rangle = |\tilde{\psi}_F\rangle \otimes |\tilde{\psi}_B\rangle$, the general form of the final state is obtained by applying the scattering operator:

$$\begin{aligned}
|\tilde{\psi}_f\rangle &= \hat{S}(t_f, t_i)|\tilde{\psi}_i\rangle \\
&= \left[cos|g| - i\, sin|g|\, \left(\hat{\sigma}^+ e^{-\frac{i\omega_{2,1}\hat{z}}{v_0}} e^{i\phi_g} + \hat{\sigma}^- e^{\frac{i\omega_{2,1}\hat{z}}{v_0}} e^{-i\phi_g}\right)\right]|\tilde{\psi}_i\rangle \\
&= \left[cos|g| - i\, sin|g|\, \left(\hat{\sigma}^+ e^{-\frac{i\omega_{2,1}\hat{z}}{v_0}} e^{i\phi_g} + \hat{\sigma}^- e^{\frac{i\omega_{2,1}\hat{z}}{v_0}} e^{-i\phi_g}\right)\right]|\tilde{\psi}_F(p)\rangle \\
&\otimes \left[sin\left(\frac{\theta}{2}\right)|1\rangle + e^{i\phi} cos\left(\frac{\theta}{2}\right)|2\rangle\right] \\
&= cos|g|\,|\tilde{\psi}_i\rangle - i\, sin|g|\, e^{i\phi_g}\left(e^{-\frac{i\omega_{2,1}\hat{z}}{v_0}}|\tilde{\psi}_F(p)\rangle\right) \otimes \left(\hat{\sigma}^+ sin\left(\frac{\theta}{2}\right)|1\rangle\right) \\
&\quad - i\, sin|g|\, e^{i\phi_g}\left(e^{\frac{i\omega_{2,1}\hat{z}}{v_0}}|\tilde{\psi}_F(p)\rangle\right) \otimes \left(\hat{\sigma}^- e^{i\phi} cos\left(\frac{\theta}{2}\right)|2\rangle\right) \\
&= cos|g|\,|\tilde{\psi}_i\rangle + |\delta\tilde{\psi}_i\rangle
\end{aligned} \tag{C1}$$

where $|\delta\tilde{\psi}_i\rangle = -i\, sin|g|\, e^{i\phi_g}|\tilde{\psi}_F(p - \delta p_{2,1})\rangle \otimes \left(sin\left(\frac{\theta}{2}\right)|2\rangle\right) - i\, sin|g|\, e^{-i\phi_g}|\tilde{\psi}_F(p + \delta p_{2,1})\rangle \otimes \left(e^{i\phi} cos\left(\frac{\theta}{2}\right)|1\rangle\right)$, $\delta p_{2,1} = \hbar\omega_{2,1}/v_0$. Since the free electron and bound electron are entangled after the interaction, it is convenient to represent the final state by the density matrix

$$\hat{\rho}_f = |\Psi_f\rangle\langle\Psi_f| = cos^2|g|\,|\Psi_i\rangle\langle\Psi_i| + cos|g|\,(|\Psi_i\rangle\langle\delta\Psi_i| + |\delta\Psi_i\rangle\langle\Psi_i|) + |\delta\Psi_i\rangle\langle\delta\Psi_i| \tag{C2}$$

where

$$\begin{aligned}
|\Psi_i\rangle\langle\delta\Psi_i| &= i\, sin|g|\, e^{i\phi_g}|\psi_p\rangle\langle\psi_{p+\delta p_{2,1}}| \otimes \left(-\frac{e^{-i\phi} sin\theta}{2}|1\rangle\langle 1| + cos^2\left(\frac{\theta}{2}\right)|2\rangle\langle 1|\right) \\
&\quad + i\, sin|g|\, e^{-i\phi_g}|\psi_p\rangle\langle\psi_{p-\delta p_{2,1}}| \otimes \left(sin^2\left(\frac{\theta}{2}\right)|1\rangle\langle 2| - \frac{e^{i\phi} sin\theta}{2}|2\rangle\langle 2|\right) \\
&= i\, sin|g| \begin{pmatrix} -\frac{e^{i(\phi_g - \phi)} sin\theta}{2}|\psi_p\rangle\langle\psi_{p+\delta p_{2,1}}| & e^{-i\phi_g} sin^2\left(\frac{\theta}{2}\right)|\psi_p\rangle\langle\psi_{p-\delta p_{2,1}}| \\ e^{i\phi_g} cos^2\left(\frac{\theta}{2}\right)|\psi_p\rangle\langle\psi_{p+\delta p_{2,1}}| & -\frac{e^{-i(\phi_g - \phi)} sin\theta}{2}|\psi_p\rangle\langle\psi_{p-\delta p_{2,1}}| \end{pmatrix}
\end{aligned} \tag{C3}$$

and

$$|\delta\Psi_i\rangle\langle\delta\Psi_i| = \sin^2|g| \begin{pmatrix} \cos^2\left(\frac{\theta}{2}\right)|\psi_{p+\delta p_{2,1}}\rangle\langle\psi_{p+\delta p_{2,1}}| & -\frac{e^{i(\phi-2\phi_g)}\sin\theta}{2}|\psi_{p+\delta p_{2,1}}\rangle\langle\psi_{p-\delta p_{2,1}}| \\ -\frac{e^{i(2\phi_g-\phi)}\sin\theta}{2}|\psi_{p-\delta p_{2,1}}\rangle\langle\psi_{p+\delta p_{2,1}}| & \sin^2\left(\frac{\theta}{2}\right)|\psi_{p-\delta p_{2,1}}\rangle\langle\psi_{p-\delta p_{2,1}}| \end{pmatrix}$$

(C4)

In order to obtain the density distribution of the free electron and the transition probability of the TLS, we partially trace the density matrix of the final state, setting $\hat{\rho}_F^f = \text{Tr}_B[\hat{\rho}_f], \hat{\rho}_B^f = \text{Tr}_F[\hat{\rho}_f]$. Consequently, the momentum density distribution of the free electron is given by:

$$\rho_p^f = Diag[\hat{\rho}_F^f] = \cos^2|g|\left|c_p^{(0)}\right|^2 -$$

$$\frac{1}{2}\sin(2|g|)\sin\theta \ \text{Re}\left\{ie^{i(\phi_g-\phi)}Co(p,\delta p_{2,1}) + ie^{-i(\phi_g-\phi)}Co(p,-\delta p_{2,1})\right\}$$ (C5)

$$\sin^2|g|\left[\cos^2\left(\frac{\theta}{2}\right)\left|c_{p+\delta p_{2,1}}^{(0)}\right|^2 + \sin^2\left(\frac{\theta}{2}\right)\left|c_{p-\delta p_{2,1}}^{(0)}\right|^2\right]$$

where $c_p^{(0)} = \langle p|\psi_p\rangle$ and $Co(p,\pm\delta p_{2,1}) = c_{p\pm\delta p_{2,1}}^{*(0)}c_p^{(0)}$ and $I(\delta p_{2,1}) = \int_p dp\, Co(p,\delta p_{2,1})$ is the auto-correlation function [42], then the final density matrix of the TLS is

$$\rho_B^f = \cos^2|g|\begin{pmatrix} \sin^2\left(\frac{\theta}{2}\right) & -\frac{e^{i\phi}\sin\theta}{2} \\ -\frac{e^{-i\phi}\sin\theta}{2} & \cos^2\left(\frac{\theta}{2}\right) \end{pmatrix}$$

$$+ \sin(2|g|)\,\text{Re}\left\{i\begin{pmatrix} -\frac{e^{i(\phi_g-\phi)}\sin\theta}{2}I(\delta p_{2,1}) & e^{-i\phi_g}\sin^2\left(\frac{\theta}{2}\right)I(-\delta p_{2,1}) \\ e^{i\phi_g}\cos^2\left(\frac{\theta}{2}\right)I(\delta p_{2,1}) & -\frac{e^{-i(\phi_g-\phi)}\sin\theta}{2}I(-\delta p_{2,1}) \end{pmatrix}\right\}$$ (C6)

$$+ \sin^2|g|\begin{pmatrix} \cos^2\left(\frac{\theta}{2}\right) & -\frac{e^{i(\phi-2\phi_g)}\sin\theta}{2}I(-2\delta p_{2,1}) \\ -\frac{e^{-i(\phi-2\phi_g)}\sin\theta}{2}I(2\delta p_{2,1}) & \sin^2\left(\frac{\theta}{2}\right) \end{pmatrix}$$

and the transition probabilities of the TLS are

$$P_1^f = \cos^2|g|\cos^2\left(\frac{\theta}{2}\right) - \frac{1}{2}\sin(2|g|)\sin\theta\,\text{Re}\left\{iI(-\delta p_{2,1})e^{-i(\phi_g-\phi)}\right\} + \sin^2|g|\sin^2\left(\frac{\theta}{2}\right)$$ (C7)

$$= P_1^i + \sin^2|g|\cos\theta - \frac{1}{2}\sin(2|g|)\sin\theta \, \text{Re}\left\{I(-\delta p_{2,1})e^{-i(\phi_g-\phi)+i\frac{\pi}{2}}\right\}$$

$$P_2^f = \cos^2|g|\sin^2\left(\frac{\theta}{2}\right) - \frac{1}{2}\sin(2|g|)\sin\theta \, \text{Re}\left\{iI(\delta p_{2,1})e^{i(\phi_g-\phi)}\right\} + \sin^2|g|\cos^2\left(\frac{\theta}{2}\right)$$

$$= P_2^i - \sin^2|g|\cos\theta - \frac{1}{2}\sin(2|g|)\sin\theta \, \text{Re}\left\{I(\delta p_{2,1})e^{i(\phi_g-\phi)+i\frac{\pi}{2}}\right\}$$

**APPENDIX D: GAUSSIAN QEW BROADENING AND CHIRP**

In momentum space, we model the free electron as a Gaussian wavepacket $|\widetilde{\psi}_F(p)\rangle = \sum_p c_p|p\rangle$, where $c_p = \frac{1}{(2\pi\sigma_p^2)^{1/4}}\exp\left[-\frac{(p-p_0)^2}{4\sigma_p^2}\right]$. The free propagation of a Gaussian-shaped QEW can be described in momentum space as: $|\widetilde{\psi}_F(p,t)\rangle = e^{-i\widehat{H}_{0F}t/\hbar}|\psi_F(p)\rangle$, which means that the electron wavefunction becomes

$$c_p \to c_p(t) = \frac{1}{(2\pi\sigma_p^2)^{1/4}} e^{-\frac{iE_pt}{\hbar}} \exp\left[-\frac{(p-p_0)^2}{4\sigma_p^2}\right] \tag{D1}$$

The additional phase term is determined by the electron energy dispersion relation $E_p = E_0 + v_0(p-p_0) + \frac{1}{2\gamma^3 m}(p-p_0)^2$. Combining the quadratic term of $(p-p_0)$ yields

$$c_p(t) = \frac{1}{(2\pi\sigma_p^2)^{1/4}} e^{-i[E_0+v_0(p-p_0)]t/\hbar} \exp\left[-\frac{(p-p_0)^2}{4\tilde{\sigma}_p^2(t)}\right] \tag{D2}$$

We define the momentum spread after propagation as

$$\tilde{\sigma}_p(t) = \sqrt{\frac{\sigma_p^2}{1 + i\frac{2t}{\gamma^3 m\hbar}\sigma_p^2}} \tag{D3}$$

Consequently, the wavefunction in coordinates domain is given by:

$$\begin{aligned}\psi_z(t) &= \int \frac{dp}{\sqrt{2\pi\hbar}} c_p(t) e^{ipz/\hbar} \\ &= \frac{1}{(2\pi\sigma_p^2)^{1/4}} e^{i(p_0 z - E_0 t)/\hbar} \int dp \exp\left[-\frac{(p-p_0)^2}{4\tilde{\sigma}_p^2(t)}\right] e^{i(p-p_0)(z-v_0 t)/\hbar} \\ &= \frac{1}{(2\pi\tilde{\sigma}_z^2(t))^{1/4}} \exp\left[-\frac{(z-v_0 t)^2}{4\tilde{\sigma}_z^2(t)}\right]\end{aligned} \tag{D4}$$

where the chirped wavepacket size is

$$\tilde{\sigma}_z(t) = \sigma_{z0}\sqrt{1 + i\frac{\hbar t}{2\gamma^3 m\sigma_{z0}^2}} \tag{D5}$$

and $\sigma_{z0} = \hbar/2\sigma_p$. When considering the density distribution $|\psi_z(t)|^2$, the wavepacket size (standard deviation of the density function envelope) is $\sigma_z(t) = |\tilde{\sigma}_z(t)| = \sigma_{z0}\sqrt{1 + \left(\frac{\hbar t}{2\gamma^3 m\sigma_{z0}^2}\right)^2}$. Thus, even if the initial wavepacket $\sigma_{t0}$ approaches 0, the real wavepakcet is always finite. According to the inequality of arithmetic and geometric means $|a| + |b| \geq 2\sqrt{|ab|}$, we can derive the lower limit of the wavepacket size for any given free drift time $t_D$:

$$\sigma_z(t_D) \geq \sigma_z(t_D)|_{min} = \sqrt{\frac{\hbar t_D}{\gamma^3 m_e}} \tag{D6}$$

We find that in order to mitigate the QEW envelop broadening effect, and satisfy the near point-particle condition at the FEBERI interaction point $\Gamma = \omega_{2,1}\sigma_t(L_D) < 1$, the QEW free drift length must be kept within the limitation $L_D < z_G$, where

$$z_G = vt_D|_{\Gamma=\sqrt{2}} = \frac{\gamma^3 m_e v_0^3}{\hbar \omega_{2,1}^2} \tag{D7}$$

**APPENDIX E: DERIVATION OF POST_INTERACTION ELECTRON MOMENTUM DISTRIBUTION AND TLS TRANSITION PROBABILTY FOR THE CESE OF A GAUSSIAN QEW**

The general form of the first-order free electron post-interaction density distribution (Eq. C5) is

$$\Delta\rho^{(1)} = \frac{1}{2}\sin(2|g|)\sin\theta \, \text{Re}\{ie^{i\phi_g - i\phi}Co(p, \delta p_{2,1}) + ie^{-i\phi_g + i\phi}Co(p, -\delta p_{2,1})\} \tag{E1}$$

With the initial electron wavefunction

$$c_p^{(0)} = c_p(t_D) = \frac{1}{(2\pi\sigma_p^2)^{1/4}} e^{-i[E_0 + v_0(p-p_0)]t_D/\hbar} \exp\left[-\frac{(p-p_0)^2}{4\tilde{\sigma}_p^2(t_D)}\right] \tag{E2}$$

where $\tilde{\sigma}_p(t_D) = \sqrt{\frac{\sigma_p^2}{1 + i\frac{2t_D}{\gamma^3 m\hbar}\sigma_p^2}}$, and $Co(p, \pm\delta p_{2,1}) = c_{p\pm\delta p_{2,1}}^{*(0)} c_p^{(0)}$, we obtain:

$$\begin{aligned}Co(p, \delta p_{2,1}) = &\frac{1}{\sqrt{2\pi\sigma_p^2}} e^{i\delta p_{2,1}t_D/\hbar} \exp\left[-\frac{(p-p_0+\delta p_{2,1})^2}{4\sigma_p^2} + i\frac{(p-p_0+\delta p_{2,1})^2 t_D}{2\gamma^3 m\hbar}\right] \\ &\times \exp\left[-\frac{(p-p_0)^2}{4\sigma_p^2} - i\frac{(p-p_0)^2 t_D}{2\gamma^3 m\hbar}\right]\end{aligned} \tag{E3}$$

$$= \frac{e^{i\delta p_{2,1} t_D/\hbar}}{\sqrt{2\pi\sigma_p^2}} \exp\left[-\frac{\left(p - p_0 + \frac{\delta p_{2,1}}{2}\right)^2}{2\sigma_p^2} - \frac{\delta p_{2,1}^2}{8\sigma_p^2} + \frac{i\delta p_{2,1} t_D}{\gamma^3 m\hbar}\left(p - p_0 + \frac{\delta p_{2,1}}{2}\right)\right]$$

$$= \frac{e^{i\delta p_{2,1} t_D/\hbar}}{\sqrt{2\pi\sigma_p^2}} e^{-\Gamma^2/2} \exp\left[-\frac{\left(p - p_0 + \frac{\delta p_{2,1}}{2}\right)^2}{2\sigma_p^2} + \frac{i\delta p_{2,1} t_D}{\gamma^3 m\hbar}\left(p - p_0 + \frac{\delta p_{2,1}}{2}\right)\right]$$

and correspondingly have

$$Co(p, -\delta p_{2,1}) = \frac{e^{-\Gamma^2/2}}{\sqrt{2\pi\sigma_p^2}} e^{-i\delta p_{2,1} t_D/\hbar} \exp\left[-\frac{\left(p - p_0 - \frac{\delta p_{2,1}}{2}\right)^2}{2\sigma_p^2} - \frac{i\delta p_{2,1} t_D}{\gamma^3 m\hbar}\left(p - p_0 - \frac{\delta p_{2,1}}{2}\right)\right]$$

(E4)

Substituting Eqs. (E3) and (E4) into the expression for the first-order incremental momentum distribution, Eq. (E1), yields

$$\Delta\rho_p^{(1)} = \frac{1}{2\sqrt{2\pi\sigma_p^2}} \sin(2|g|) \sin\theta\, e^{-\frac{\Gamma^2}{2}} \times$$

$$\left\{-\sin\left[\omega_{2,1} t_D\left(1 + \frac{p - p_0 + \frac{\delta p_{2,1}}{2}}{\gamma^3 m v_0}\right) + \phi_g - \phi\right] \exp\left[-\frac{\left(p - p_0 + \frac{\delta p_{2,1}}{2}\right)^2}{2\sigma_p^2}\right]\right.$$

$$\left. + \sin\left[\omega_{2,1} t_D\left(1 + \frac{p - p_0 - \frac{\delta p_{2,1}}{2}}{\gamma^3 m v_0}\right) + \phi_g - \phi\right] \exp\left[-\frac{\left(p - p_0 - \frac{\delta p_{2,1}}{2}\right)^2}{2\sigma_p^2}\right]\right\}$$

$$\simeq \frac{1}{2\sqrt{2\pi\sigma_p^2}} \sin(2|g|) \sin\theta\, e^{-\frac{\Gamma^2}{2}} \sin(\xi + \phi_g) \left\{e^{-\frac{\left(p - p_0 - \frac{\delta p_{2,1}}{2}\right)^2}{2\sigma_p^2}} - e^{-\frac{\left(p - p_0 + \frac{\delta p_{2,1}}{2}\right)^2}{2\sigma_p^2}}\right\}$$

(E5)

where $\zeta = \omega_{2,1} t_D - \phi$.

For derivation of the FEBERI induced incremental occupation probability of the upper and lower levels of the TLS ($\Delta P_2 = -\Delta P_1$) we need to evaluate first the auto-correlation function $I(\delta p_{2,1}) = \int_p dp\, Co(p, \delta p_{2,1})$. Inserting Eq. (E4) into $I(\delta p)$ results in

$$I(\delta p_{2,1}) = \frac{1}{\sqrt{2\pi\sigma_p^2}} \int dp\, e^{i(E_{p+\delta p_{2,1}} - E_p) t_D/\hbar} \exp\left[-\frac{(p - p_0 + \delta p_{2,1})^2}{4\sigma_p} - \frac{(p - p_0)^2}{4\sigma_p}\right] \quad (E6)$$

$$= \frac{e^{i\omega_{2,1}t_D}e^{-\Gamma^2/2}}{\sqrt{2\pi\sigma_p^2}} \int dp \, e^{i\omega_{2,1}t_D \frac{\left(p-p_0+\frac{\delta p_{2,1}}{2}\right)}{\gamma^3 m v_0}} exp\left[-\frac{\left(p-p_0+\frac{\delta p_{2,1}}{2}\right)^2}{2\sigma_p^2}\right]$$

$$= e^{i\omega_{2,1}t_D}e^{-(\Gamma^2+\Gamma_D^2)/2}$$

where the envelop decay factor due to energy dispersion and drift is defined as

$$\Gamma_D = \omega_{2,1}L_D \frac{\sigma_E}{\gamma^3 m_e v_0^3} = \frac{1}{2\omega_{2,1}\sigma_{t0}} \frac{L_D}{z_G} = \frac{\sigma_E}{E_{2,1}} \frac{L_D}{z_G} \tag{E7}$$

We finally get

$$\Delta P_2 = -\sin^2|g|\cos\theta - \frac{1}{2}\sin(2|g|)\sin\theta \sin(\omega_{2,1}t_D - \phi + \phi_g) e^{-(\Gamma^2+\Gamma_D^2)/2} \tag{E8}$$

The numerically computed incremental electron energy distribution $\Delta\rho_{E_p}$ is depicted in Fig. 2 of the main text. The continuous parameters dependence version of the cases shown in Fig. 2 are presented in movie: (S1) the dependence on $\phi$ in the case of a short QEW; (S2) the dependence on $\theta$ in the case of a long QEW; (S3) the dependence on both $\phi$ and $\theta$ in an intermediate case:

# APPENDIX F: DETAILED ANALYSIS OF DENSITY BUNCHING

In this section, we derive the bunched density distribution $\rho(z,t)$ of the QEW after energy modulation and drift. We start with the energy modulation of the QEW modeled by a Gaussian distribution in momentum domain, as given in Appendix D, the QEW can be written as $|\psi_i(p)\rangle = \sum_p c_p|p\rangle$, where $c_p^{(0)} = \frac{1}{(2\pi\sigma_p^2)^{1/4}} \exp\left[-\frac{(p-p_0)^2}{4\sigma_p^2}\right]$. The energy modulation can be described by the scattering operator

$$\hat{S} = \exp\left[-\frac{i}{\hbar}\int_{t_0-\frac{t_I}{2}}^{t_0+\frac{t_I}{2}} dt \, \hat{H}_{int}(t)\right] \tag{F1}$$

where $\hat{H}_{int}(t) = \frac{e}{\gamma m}\hat{A}(z,t) \cdot \hat{p}$ describes the coupling between the free electron and the laser-induced optical near-field, $t_0$ represent the the arrival time of the QEW centroid to the center of the interaction region and $t_I$ is the interaction time. For simplicity we consider here the case of PINEM modulation in a case of "transition radiative interaction, namely the case where the spatial extent of the near field of a laser milluminated nanostructure (e.g. a tip) is shorter than the optical wavelength of the laser [19]. In this case the QEW hardly senses the spatial variation of the time-varying near field during the interaction, and the vector potential can be generally written as $\hat{A}(z,t) = A_0 \sin\left[\omega_L\left(\frac{\hat{z}}{v_0} - t\right) + \phi_i\right]$, where $\phi_i$ is the initial phase of the near-field at the interaction time $t_0$. Since the initial momentum distribution of the QEW (Eq. D1) is narrow, $\sigma_p \ll \hbar\frac{\omega_L}{v_0}$, it is valid to assume $\hat{p}|p\rangle \approx p_0|p\rangle$. Thus, the explicit form of scattering operator is

$$\hat{S} = \exp\left[-\frac{i}{\hbar}\int_{t_0-\frac{t_I}{2}}^{t_0+\frac{t_I}{2}} dt\, \hat{H}_{int}(t)\right] = \exp\left[-\frac{ieA_0 p_0}{\gamma m\hbar}\int_{t_0-\frac{t_I}{2}}^{t_0+\frac{t_I}{2}} dt\, \sin\left[\omega_L\left(\frac{\hat{z}}{v_0} - t\right) + \phi_i\right]\right]$$
$$= \exp\left[-i\frac{2eA_0 p_0}{\gamma m\hbar}\sin(\omega_L t_I)\sin\left(\frac{\omega_L}{v_0}\hat{z} + \phi_i - \omega_L t_0\right)\right] \quad (F2)$$
$$= \exp\left[-i2g_L \sin\left(\frac{\omega_L}{v_0}\hat{z} + \phi_i - \omega_L t_0\right)\right]$$

where $\phi_0 = \omega_L t_0 - \phi_i$ and the coupling constant $g_L = \frac{eA_0 p_0}{\gamma m\hbar}\sin(\omega_L t_I)$.

According to the Jacobi-Anger expansion, the scattering matrix can be presented in terms of a serries:

$$\hat{S} = \sum_{n=-\infty}^{+\infty} J_n(2|g|)\exp[-in(\delta k_L \hat{z} - \phi_0)] \quad (F3)$$

Thus, the energy-modulated QEW has the form

$$|\psi(p)\rangle = \hat{S}|\psi_i(p)\rangle = \sum_{m=-\infty}^{+\infty} J_m(2|g|)\exp[im(\delta k_L \hat{z} - \phi_0)]\sum_p c_p^{(0)}|p\rangle$$
$$= \sum_p c_p^{(0)}\sum_{m=-\infty}^{+\infty} J_m(2|g|)\exp[im(\delta k_L \hat{z} - \phi_0)]|p\rangle$$
$$= \sum_p c_p^{(0)}\sum_{m=-\infty}^{+\infty} J_m(2|g|)\exp(-im\phi_0)|p + m\hbar\delta k_L\rangle \quad (F3)$$
$$= \sum_p c_{p-m\hbar\delta k_L}^{(0)}\sum_{m=-\infty}^{+\infty} J_m(2|g|)\exp(-im\phi_0)|p\rangle$$
$$= \sum_p c_p|p\rangle$$

where the new coefficient $c_p$ of the energy-modulated QEW has the form

$$c_p = \frac{1}{(2\pi\sigma_p^2)^{\frac{1}{4}}}\sum_{m=-\infty}^{+\infty} J_m(2|g|)\exp\left[-\frac{(p-p_0-m\delta p_L)^2}{4\sigma_p^2}\right]\exp(-im\phi_0) \quad (F4)$$

where $\delta p_L = \hbar\omega_L/v_0$. Thus, the momentum presentation of the wavefuction after drift:

$$c_p(t_D) = \frac{e^{-iE_p t_D}}{(2\pi\sigma_p^2)^{\frac{1}{4}}}\sum_m J_m(2|g_L|)e^{-\frac{(p-p_0-m\delta p_L)^2}{4\sigma_p^2}-im\phi_0} = e^{-iE_p t_D}\sum_m c_p^m \quad (F5)$$

where $c_p^m = \frac{1}{(2\pi\sigma_p^2)^{1/4}} J_m(2|g_L|)e^{-\frac{(p-p_0-m\delta p_L)^2}{4\sigma_p^2}-im\phi_0}$. The wavefunction in coordinate space is derived by Fourier transformation $\psi(z, t_D) = \frac{1}{\sqrt{2\pi\hbar}}\int dp\, c_p e^{ipz/\hbar}$. Because this is a linear operation, the coordinate presentation of the wavefunction is also a sum of satellite wavepackets $\psi(z,t_D) = \sum_m \psi_m(z,t_D) = \sum_m \frac{1}{\sqrt{2\pi\hbar}} \int dp \, c_p^m e^{i(pz-E_p t_D)/\hbar}$, where,

$$\begin{aligned}
\psi_m(z,t_D) &= \frac{1}{\sqrt{2\pi\hbar}} J_m(2|g_L|) e^{-im\phi_0} \frac{1}{(2\pi\sigma_p^2)^{1/4}} \int dp \, e^{-\frac{(p-p_0-m\delta p_L)^2}{4\sigma_p^2}} e^{i(pz-E_p t_D)/\hbar} \\
&= \frac{J_m(2|g_L|)}{\sqrt{2\pi\hbar}} e^{-im\phi_0} \frac{e^{i(p_0 z - E_0 t_D)/\hbar}}{(2\pi\sigma_p^2)^{1/4}} \int dp \, e^{-\frac{(p-m\delta p_L)^2}{4\sigma_p^2}} e^{-i\frac{t_D}{2\gamma^3 m_e \hbar} p^2} e^{ip(z-v_0 t_D)/\hbar} \\
&= \frac{J_m(2|g_L|)}{\sqrt{2\pi\hbar}} e^{-im\phi_0} \frac{e^{\frac{i(p_0 z - E_0 t_D)}{\hbar}}}{(2\pi\sigma_p^2)^{\frac{1}{4}}} \times
\end{aligned}$$

$$\int dp \exp\left[-\left(\frac{1}{4\sigma_p^2} + \frac{it_D}{2\gamma^3 m_e \hbar}\right) p^2 + \frac{m\delta p_L}{2\sigma_p^2} p - \frac{m^2 \delta p_L^2}{4\sigma_p^2}\right] e^{ip(z-v_0 t_D)/\hbar}$$

$$= \frac{J_m(2|g_L|)}{\sqrt{2\pi\hbar}} e^{-im\phi_0} \frac{e^{\frac{i(p_0 z - E_0 t_D)}{\hbar}}}{(2\pi\sigma_p^2)^{1/4}} \exp\left[\frac{m^2 \delta p_L^2}{4\sigma_p^2}\left(\frac{\tilde{\sigma}_p^2(t_D)}{\sigma_p^2} - 1\right)\right]$$

$$\times \int dp \exp\left[-\frac{1}{4\tilde{\sigma}_p^2}\left(p - \frac{\tilde{\sigma}_p^2(t_D)}{\sigma_p^2} m\delta p_L\right)^2\right] e^{ip(z-v_0 t_D)/\hbar}$$

(F6)

where $\tilde{\sigma}_p^2(t_D) = \sigma_p^2 / \left(1 + \frac{2it_D}{\gamma^3 m_e \hbar} \sigma_p^2\right)$ is the same as Eq. (D3), which using the standard Gaussian integral results in

$$\psi_m(z,t_D) = \frac{e^{\frac{i(p_0 z - E_0 t_D)}{\hbar}}}{(2\pi\sigma_z^2(t_D))^{\frac{1}{4}}} J_m(2|g_L|) e^{-im\phi_0} \exp\left[-\frac{(z-v_0 t_D)^2}{\sigma_z^2(t_D)}\right]$$

$$\times \exp\left[\frac{m^2 \delta p_L^2}{4\sigma_p^2}\left(\frac{\tilde{\sigma}_p^2(t_D)}{\sigma_p^2} - 1\right)\right] e^{im\frac{\tilde{\sigma}_p^2(t_D)\delta p_L(z-v_0 t_D)}{\sigma_p^2 \hbar}}$$

(F7)

where $\tilde{\sigma}_z(t_D) = \sigma_z \sqrt{1 + \frac{i\hbar t_D}{2\gamma^3 m_e \sigma_{z0}^2}}$. Since in the case of a modulated QEW the envelop must be broad, necessarily $\sigma_p = \hbar/2\sigma_{z_0} \to 0$, so the chirp effect is ignorable and $\tilde{\sigma}_z(t_D) \simeq \sigma_{z_0}$. We get

$$\frac{\tilde{\sigma}_p^2(t_D)}{\sigma_p^2} = \frac{1}{1 + \frac{2it_D}{\gamma^3 m_e \hbar} \sigma_p^2} = \frac{1 - \frac{2it_D}{\gamma^3 m_e \hbar} \sigma_p^2}{1 + \left(\frac{2t_D \sigma_p^2}{\gamma^3 m_e \hbar}\right)^2} \simeq 1 - \frac{2it_D}{\gamma^3 m_e \hbar} \sigma_p^2$$

After neglecting high orders of $\sigma_p$, we find that the spatial wavefunction $\psi(z,t_D) = \sum_m \psi_m(z,t_D)$ can be written as:

$$\psi(z,t_D) = \frac{e^{\frac{i(p_0 z - E_0 t_D)}{\hbar}}}{(2\pi\sigma_{z0}^2)^{\frac{1}{4}}} \exp\left[-\frac{z(t)^2}{4\sigma_{z0}^2}\right] \sum_m J_m(2|g_L|) e^{-im\phi_0} e^{im\delta p_L \left(\frac{z(t)}{\hbar} - \frac{t_D m \delta p_L}{2\gamma^3 m_e \hbar}\right)} \tag{F8}$$

where $z(t) = z - v_0 t_D$. Thus, the density distribution of the modulated electron is

$$\begin{aligned}
\rho(z,t_D) &= |\psi(z,t_D)|^2 \\
&= \frac{\exp\left[-\frac{z(t)^2}{2\sigma_{z0}^2}\right]}{\sqrt{2\pi\sigma_{z0}^2}} \sum_{m,n} J_m^*(2|g_L|) J_n(2|g_L|) e^{i(m-n)\phi_0} \exp\left[i(n-m)\frac{\delta p_L z(t)}{\hbar}\right] \\
&\quad \times \exp\left[i(m^2-n^2)\frac{\delta p_L^2 \chi t_D}{4\sigma_p^2}\right] \\
&= \sum_l e^{il\phi_0 - il\delta p_L z(t)/\hbar} \exp\left[2\pi i \frac{l^2 L_D}{z_T}\right] \sum_n J_{n+l}(2|g_L|) J_n(2|g_L|) e^{4\pi i n \frac{l L_D}{z_T}}
\end{aligned} \tag{F9}$$

where $z_T = \frac{4\pi\gamma^3 m_e v_0^3}{\hbar \omega_b^2}$ is the Talbot length and $L_D = v_0 t_D$ is the free drift propagation length. According to Graf's addition theorem, the density distribution is

$$\rho_z = \frac{\exp\left[-\frac{z(t)^2}{2\sigma_z^2(t_D)}\right]}{\sqrt{2\pi\sigma_z^2(t_D)}} \sum_m e^{im\phi_0 - \frac{im\delta p_L z(t)}{\hbar}} \exp\left[2\pi i \frac{m^2 L_D}{z_T}\right] J_m\left[4|g_L|\sin\left(2\pi m \frac{L_D}{z_T}\right)\right] \tag{F10}$$

The evolution of the density distribution of a modulated QEW, corresponding to Fig. 4, is shown in the Movie S4.

## APPENDIX G: DERIVATION OF POST_INTERACTION MOMENTUM DISTRIBUTION AND TLS TRANSITIOIN PROBABILITY FOR A MODULETED QEW

In section S. C, we derived a generic expression (Eq. C5) for the post-interaction density distribution of the free electron $\rho_p^f$. For the case of a Gaussian QEW that is momentum-modulated by a PINEM process, and then traverses a drift-time $t_D$, we plug in there the initial state in momentum presentation $c_p^{(0)} = c_p(t_D) = \frac{1}{(2\pi\sigma_p^2)^{1/4}} e^{-iE_p t_D/\hbar} \sum_m J_m(2|g_L|) \exp\left[-\frac{(p - p_0 - m\delta p_L)^2}{4\sigma_p^2} + im\phi_0\right]$, where $\phi_0$ is the initial phase of the modulation laser and $\delta p_L = \hbar\omega_b/v_0$, $\omega_b$ is the frequency of the modulating laser. With this wavefunction, Eq. (C5) results in the following expression for the post-interaction density distribution of the free electron:

$$\begin{aligned}
\rho^f(p) &= \cos^2|g| \, |c_p(t_D)|^2 + \sin^2|g| \left[\cos^2\left(\frac{\theta}{2}\right) |c_{p+\delta p_{2,1}}(t_D)|^2 + \sin^2\left(\frac{\theta}{2}\right) |c_{p-\delta p_{2,1}}(t_D)|^2\right] \\
&\quad -\frac{1}{2}\sin(2|g|)\sin\theta \, \mathrm{Re}\{ie^{i\phi_g - i\phi} Co(p, \delta p_{2,1}) + ie^{-i\phi_g + i\phi} Co(p, -\delta p_{2,1})\} \\
&= \rho^{(i)}(p) + \Delta\rho(p) = \rho^{(i)}(p) + \Delta\rho^{(0)}(p) + \Delta\rho^{(1)}(p) + \Delta\rho^{(2)}(p)
\end{aligned}$$

(G1)

where $\delta p_{2,1} = \hbar\omega_{2,1}/v_0$.

We first calculate the zero order and second order incremental density distribution terms in (Eq. G1). Since the quantum recoil induced by the modulating laser is much larger than the momentum spread of the QEW $\frac{\hbar\omega_L}{v_0} \gg \sigma_p$, the overlap between the sidebands is negligible, and we can make the approximation:

$\left|\sum_m J_m(2|g_L|)\exp\left[-\frac{(p-p_0-m\delta p_L)^2}{4\sigma_p^2} + in\phi_0\right]\right|^2 \simeq \sum_m |J_m(2|g_L|)|^2 \exp\left[-\frac{(p-p_0-m\delta p_L)^2}{2\sigma_p^2}\right]$, and we get

$$\Delta\rho_p^{(0)} = -\frac{\sin^2|g|}{\sqrt{2\pi\sigma_p^2}} \sum_m |J_m(2|g_L|)|^2 \exp\left[-\frac{(p-p_0-m\delta p_L)^2}{2\sigma_p^2}\right] \tag{G2}$$

$$\Delta\rho_p^{(2)} = \frac{\sin^2|g|}{\sqrt{2\pi\sigma_p^2}} \sum_m \left[|J_m(2|g_L|)|^2 \cos^2\left(\frac{\theta}{2}\right) e^{-\frac{(p-p_0-m\delta p_L+\delta p_{2,1})^2}{2\sigma_p^2}} \right.$$
$$\left. + |J_m(2|g_L|)|^2 \sin^2\left(\frac{\theta}{2}\right) e^{-\frac{(p-p_0-m\delta p_L-\delta p_{2,1})^2}{2\sigma_p^2}}\right] \tag{G3}$$

$$\simeq \frac{\sin^2|g|}{\sqrt{2\pi\sigma_p^2}} \sum_m B_{n,m}(\theta) \exp\left[-\frac{(p-p_0-m\delta p_L)^2}{2\sigma_p^2}\right]$$

where $B_{n,m}(\theta) = |J_{m-n}(2|g_L|)|^2 \cos^2\left(\frac{\theta}{2}\right) + |J_{m+n}(2|g_L|)|^2 \sin^2\left(\frac{\theta}{2}\right)$.

In order to evaluate the first-order term in (Eq. G1), we need first to evaluate the function $Co(p, \delta p_{2,1})$:

$$Co(p, \delta p_{2,1}) = \frac{1}{\sqrt{2\pi\sigma_p^2}} e^{i(E_{p+\delta p_{2,1}} - E_p)t_D/\hbar} \sum_{n,m} J_n(2|g_L|) J_m(2|g_L|) \times$$
$$\exp\left[-\frac{(p-p_0+\delta p_{2,1}-n\delta p_L)^2}{4\sigma_p^2} - \frac{(p-p_0-m\delta p_L)^2}{4\sigma_p^2} - i(n-m)\phi_0\right]$$

$$= \frac{e^{i\omega_{2,1}t_D\left(1+\frac{p-p_0+\frac{\delta p_{2,1}}{2}}{\gamma^3 m_e v_0}\right)}}{\sqrt{2\pi\sigma_p^2}} \sum_{n,m} J_n(2|g_L|) J_m(2|g_L|) e^{-i(n-m)\phi_0} \times$$
$$\exp\left[-\frac{\left(p-p_0+\frac{\delta p_{2,1}}{2}-\frac{m+n}{2}\delta p_L\right)^2}{2\sigma_p^2}\right]\exp\left[-\frac{(\delta p_{2,1}-(n-m)\delta p_L)^2}{8\sigma_p^2}\right] \tag{G4}$$

$$= \frac{e^{i\omega_{2,1}t_D\left(1+\frac{p-p_0+\frac{\delta p_{2,1}}{2}}{\gamma^3 m_e v_0}\right)}}{\sqrt{2\pi\sigma_p^2}} \sum_{m,l} J_{m+l}(2|g_L|) J_m(2|g_L|) e^{il\phi_0 - \frac{(\omega_{2,1}-l\omega_b)^2 \sigma_{t0}^2}{2}}$$

$$\times \exp\left[-\frac{(p - p_0 - m\delta p_L + (\delta p_{2,1} - l\delta p_L)/2)^2}{2\sigma_p^2}\right]$$

Considering the nearly resonant case,

$$\omega_{2,1} \approx n\omega_b \ (\delta p_{2,1} \approx n\delta p_L) \tag{G5}$$

we neglect all terms except $l = n$:

$$Co(p, \delta p_{2,1}) \simeq \frac{1}{\sqrt{2\pi\sigma_p^2}} e^{i\omega_{2,1}t_D\left(1 + \frac{p - p_0 + \frac{\delta p_{2,1}}{2}}{\gamma^3 m_e v_0}\right) + in\phi_0} e^{-\frac{(\omega_{2,1} - n\omega_b)^2 \sigma_{t0}^2}{2}}$$

$$\times \sum_m J_{m+n}(2|g_L|) J_m(2|g_L|) \exp\left[-\frac{(p - p_0 - m\delta p_L)^2}{2\sigma_p^2}\right] \tag{G6}$$

Similarly:

$$Co(p, -\delta p_{2,1}) \simeq \frac{1}{\sqrt{2\pi\sigma_p^2}} e^{-i\omega_{2,1}t_D\left(1 + \frac{p - p_0 - \frac{\delta p_{2,1}}{2}}{\gamma^3 m_e v_0}\right) - in\phi_0} e^{-\frac{(\omega_{2,1} - n\omega_b)^2 \sigma_{t0}^2}{2}}$$

$$\times \sum_m J_{m-n}(2|g_L|) J_m(2|g_L|) \exp\left[-\frac{(p - p_0 - m\delta p_L)^2}{2\sigma_p^2}\right] \tag{G7}$$

Therefore, we can write the first order term of the incremental final sideband's spectrum (Eq. G1) for a modulated Gaussian QEW in a compact form:

$$\Delta\rho_p^{(1)} = \frac{1}{2}\sin(2|g|)\sin\theta \ \text{Re}\{ie^{i\phi_g - i\phi} Co(p, \delta p_{2,1}) + ie^{-i\phi_g + i\phi} Co(p, -\delta_{2,1})\}$$

$$= \frac{1}{2}\sin(2|g|)\sin\theta \ e^{-\frac{(\omega_{2,1} - n\omega_b)^2 \sigma_{t0}^2}{2}} \sum_m A_{n,m}(\phi) \frac{1}{\sqrt{2\pi\sigma_p^2}} \exp\left[-\frac{(p - p_0 - m\delta p_L)^2}{2\sigma_p^2}\right] \tag{G8}$$

where the amplitude function $A_{n,m}(\phi)$ is

$$A_{n,m}(\phi) = \left[-\sin(\zeta + \phi_g + \phi_{\delta p_L}^m) J_{m+n}(2|g_L|) \right.$$
$$\left. + \sin(\zeta + \phi_g + \phi_{-\delta p_L}^m) J_{m-n}(2|g_L|)\right] J_m(2|g_L|) \tag{G9}$$

where $\zeta = \omega_{2,1}t_D - \phi + n\phi_0$ represents the phase of the TLS relative to the laser-modulated QEW and its arrival phase at the interaction point. The additional phase $\phi_{\pm\delta p_L} = \omega_{2,1}t_D \frac{p - p_0 \pm \frac{\delta p_{2,1}}{2}}{\gamma^3 m_e v_0}$ is a recoil-dependent (positive or negative) phase shift accumulated in drift. Under the near resonance condition Eq. (G5) and with narrow sideband width relative to the recoil momentum - $\sigma_p \ll \delta p_L$, we can set $\delta p_{2,1} \approx n\delta p_L$, and $p - p_0 = m\delta p_L$, and define the phase shift $\phi^m_{\delta p_{2,1}}$ of the m-th order sideband:

$$\phi^m_{\pm \delta p_L} = \omega_{2,1} t_D \frac{m\delta p_L \pm \frac{\delta p_{2,1}}{2}}{\gamma^3 m_e v_0} \simeq \omega_{2,1} t_D \frac{(m \pm n/2)\delta p_L}{\gamma^3 m_e v_0} \tag{G10}$$

We examine the case of no drift $t_D = 0$ (that corresponds to no density bunching of the QEW). In this case $\phi^m_{\pm \delta p_L} = 0$, and the coefficient (Eq. G9) reduces to $A_{n,m} = \sin(\zeta + \phi_g)[-J_{m+n}(2|g_L|) + J_{m-n}(2|g_L|)]J_m(2|g_L|)$. This coefficient satisfies a symmetry relation:

$$\begin{aligned}
A_{n,-m} &= \sin(\zeta + \phi_g)\left[J_{-m-n}(2|g_L|)J_{-m}(2|g_L|) - J_{-m+n}(2|g_L|)J_{-m}(2|g_L|)\right] \\
&= \sin(\zeta + \phi_g)\left[(-1)^{m+n}(-1)^m J_{m+n}(2|g_L|)J_m(2|g_L|) \right. \\
&\qquad \left. - (-1)^{m-n}(-1)^m J_{m-n}(2|g_L|)J_m(2|g_L|)\right] \\
&= \sin(\zeta + \phi_g)\left[(-1)^n J_{m+n}(2|g_L|)J_m(2|g_L|) - (-1)^n J_{m-n}(2|g_L|)J_m(2|g_L|)\right] \\
&= (-1)^{n+1} A_{n,m}
\end{aligned}$$

Thus, for $n$ being odd, the final incremental spectrum would be symmetric, while for $n$ being even, it is anti-symmetric.

We proceed now to calculate the incremental transition probability of the TLS that corresponds to the incremental distribution of the modulated QEW. For this we need first to evaluate the autocorrelation function for the modulated Gaussian case:

$$\begin{aligned}
I(\delta p_{2,1}) &= \int dp\, Co(p, \delta p_{2,1}) \\
&= \frac{1}{\sqrt{2\pi\sigma_p^2}} e^{i\omega_{2,1} t_D} \int dp\, e^{i\omega_{2,1} t_D \frac{\left(p - p_0 + \frac{\delta p_{2,1}}{2}\right)}{\gamma^3 m_e v_0}} \sum_{n,m} J_n(2|g_L|)J_m(2|g_L|)\, e^{-i(n-m)\phi_0} \times \\
&\qquad \exp\left[-\frac{\left(p - p_0 + \frac{\delta p_{2,1}}{2} - \frac{m+n}{2}\delta p_L\right)^2}{2\sigma_p^2}\right] \exp\left[-\frac{(\delta p_{2,1} - (n-m)\delta p_L)^2}{8\sigma_p^2}\right] \\
&= \frac{e^{i\omega_{2,1} t_D}}{\sqrt{2\pi\sigma_p^2}} \sum_{n,m} J_n(2|g_L|)J_m(2|g_L|)\, e^{-i(n-m)\phi_0} \exp\left[-\frac{(\delta p_{2,1} - (n-m)\delta p_L)^2}{8\sigma_p^2}\right] \times \\
&\qquad \int dp\, e^{i\omega_{2,1} t_D \frac{\left(p - p_0 + \frac{\delta p_{2,1}}{2}\right)}{\gamma^3 m_e v_0}} \exp\left[-\frac{\left(p - p_0 + \frac{\delta p_{2,1}}{2} - \frac{m+n}{2}\delta p_L\right)^2}{2\sigma_p^2}\right] \\
&= e^{i\omega_{2,1} t_D} e^{-\Gamma_D^2/2} \sum_{l,m} J_{m+l}(2|g_L|)J_m(2|g_L|)\, e^{-il\phi_0} \exp\left[-\frac{(\omega_{2,1} - l\omega_b)^2 \sigma_{t0}^2}{2}\right] e^{i\frac{\omega_{2,1} t_D}{2\gamma^3 m_e v_0}(2m+l)\delta p_L} \\
&= e^{i\omega_{2,1} t_D} e^{-\Gamma_D^2/2} \sum_l e^{-il\phi_0} \exp\left[-\frac{(\omega_{2,1} - l\omega_b)^2 \sigma_{t0}^2}{2}\right] e^{i\frac{\omega_{2,1} t_D}{2\gamma^3 m_e v_0} l\delta p_L} \times \\
&\qquad \sum_m J_{m+l}(2|g_L|)J_m(2|g_L|) e^{i\frac{\omega_{2,1} t_D}{\gamma^3 m_e v_0} m\delta p_L}
\end{aligned}$$

With Graf's addition theorem, one gets

$$I(\delta p) = e^{i\omega_{2,1} t_D} \, e^{-\Gamma_D^2/2} \sum_l J_l\left(4|g_L|\sin\left(2\pi n \frac{L_D}{z_T}\right)\right) e^{-il\phi_0}$$
$$\times \exp\left[-\frac{(\omega_{2,1}-l\omega_b)^2 \sigma_{t0}^2}{2}\right] e^{i\frac{\omega_{2,1} t_D}{2\gamma^3 m_e v_0} l\delta p_L} \tag{G11}$$

which yields the explicit expression for the incremental post-interaction occupation probability of the TLS quantum levels:

$$\Delta P_2 = -\Delta P_1 = -\sin^2|g|\cos\theta - \frac{1}{2}\sin(2|g|)\sin\theta \, e^{-\Gamma_D^2/2}$$
$$\times \sum_l J_l\left(4|g_L|\sin\left(2\pi n\frac{L_D}{z_T}\right)\right) \exp\left[-\frac{(\omega_{2,1}-l\omega_b)^2\sigma_{t0}^2}{2}\right] \sin\left(\zeta + \frac{\omega_{2,1} t_D}{2\gamma^3 m_e v_0} l\delta p_L + \phi_g\right) \tag{G12}$$

Assuming $\omega_{2,1} \approx n\omega_b$ (resonance with harmonic n (Eq. 2)), this can be written:

$$\Delta P_2 = -\Delta P_1 = -\sin^2|g|\cos\theta - \frac{1}{2}\sin(2|g|)\sin\theta \sin\left(\zeta + \phi_g + 2\pi n\frac{L_D}{z_T}\right)$$
$$\times J_n\left(4|g_L|\sin\left(2\pi n\frac{L_D}{z_T}\right)\right) \exp\left[-\frac{(\omega_{2,1}-n\omega_b)^2\sigma_{t0}^2}{2}\right] e^{-\Gamma_D^2/2} \tag{G13}$$

**APPENDIX H: DERIVATION OF THE INTEGRO_DIFFERENTIAL EQUATION FOR NUMERICAL SIMULATION**

In this section, we present the algorithm used for the numerical solution of an integro-differential equation that we derive from the interaction picture Schrodinger equation (Eq. 7). The pre-interaction joint wavefunction $|\widetilde{\Psi}(p,t)\rangle = |\tilde{\psi}_F(t)\rangle \otimes |\tilde{\psi}_B\rangle$ is described using the base vector set $\{|p\rangle \otimes |i\rangle\}$ which is annotated $\{|p,i\rangle\}$, where we used the definition $|\widetilde{\Psi}(p,t)\rangle = |\tilde{\psi}_F(t)\rangle \otimes |\tilde{\psi}_B\rangle = \sum_p c_p^{(0)}(t)|p\rangle \otimes \sum_i C_i |i\rangle$ (in the previous sections we set $C_1 = \sin\frac{\theta}{2}, C_2 = e^{i\phi}\cos\frac{\theta}{2}$).

Multiplying Eq. 7 by $\langle p,i|$ yields:

$$i\hbar \frac{\partial}{\partial t}\langle p,i|\widetilde{\Psi}(p,t)\rangle = \langle p,i|\widehat{\mathcal{H}}_I(t)|\widetilde{\Psi}(p,t)\rangle \tag{H1}$$

Next, the unity operator is inserted into the right-hand side of Eq. (H1):

$$i\hbar \frac{\partial}{\partial t}\langle p,i|\widetilde{\Psi}(p,t)\rangle = \int dp' \sum_j \langle p,i|\widehat{\mathcal{H}}_I(t)|p',j\rangle \langle p',j|\widetilde{\Psi}(p,t)\rangle \tag{H2}$$

where the kernel is

$$\langle p,i|\widehat{\mathcal{H}}_I(t)|p',j\rangle = \langle p,i|e^{i\widehat{H}_0 t/\hbar}\widehat{H}_I e^{-i\widehat{H}_0 t/\hbar}|p',j\rangle$$
$$= \langle p,i|\widehat{H}_I|p',j\rangle e^{i(E_p-E_{p'})t/\hbar} e^{i(E_i-E_j)t/\hbar} \tag{H3}$$

With the explicit form of the interaction Hamiltonian $\widehat{H}_I = e\hat{r}' \cdot \boldsymbol{E}(\hat{z}, r_{\perp 0})$, one obtains

$$\langle p,i|\widehat{H}_I|p',j\rangle = \langle i|e\hat{r}'|j\rangle \cdot \langle p|\boldsymbol{E}(\hat{z}, r_{\perp 0})|p'\rangle = \langle p|\boldsymbol{\mu}_{i,j} \cdot \boldsymbol{E}(\hat{z}, r_{\perp 0})|p'\rangle \tag{H4}$$

$$= \int dz \int dz' \langle p|z\rangle\langle z|\boldsymbol{\mu}_{i,j} \cdot \boldsymbol{E}(\hat{z},r_{\perp 0})|z'\rangle\langle z'|p'\rangle$$

$$= \frac{1}{2\pi\hbar} \int dz\, \boldsymbol{\mu}_{i,j} \cdot \boldsymbol{E}(z,r_{\perp 0})\, e^{-i(p-p')z/\hbar}$$

We denote the result of Eq.(H4) as $\widetilde{M}_{i,j}(p-p')$, and define $\langle p,i|\widetilde{\Psi}(p,t)\rangle \equiv C_{i,p}(t)$, where $C_{i,p}(t)$ is the probability amplitude of the entangled state $|i,p\rangle$. Since $\widetilde{M}_{i,j}(p-p') = 0$ for $i = j$, Eq. (H4) can be expressed as

$$\dot{C}_{i,p}(t) = \frac{1}{2\pi i\hbar^2}\int dp\, \widetilde{M}_{i,j}(p-p')C_{j,p'}(t)e^{-i(E_{p'}-E_p+E_{i,j})t/\hbar} \tag{H5}$$

where $E_{i,j} = E_i - E_j$, and $i \neq j$. Eq. (H5) can be solved numerically using Euler method by discretizing it over a finite momentum range $(-P_{\text{cutoff}}, P_{\text{cutoff}})$ with N points sampling. The cutoff of the momentum must satisfy $P_{cutoff} > \hbar\omega_{2,1}/v_0$. The discrete version of (Eq. H5) is:

$$\dot{C}_{i,p_m} = \frac{1}{2\pi i\hbar^2}\sum_{n=1}^{N} \Delta p\, \widetilde{M}_{i,j}(p_m-p_n)\, C_{j,p_n}(t)\, e^{-i(E_{p_n}-E_{p_m}+E_{i,j})t/\hbar}$$

where $p_n = \left(1-\frac{2n}{N}\right)P_{\text{cutoff}}$ and $E_{p_n} = \varepsilon_0 + v_0(p_n-p_0) + \frac{(p_n-p_0)^2}{2\gamma^3 m}$. Thus, the probability amplitude can be written as a discrete vector:

$$v_i(t) = \left[C_{i,p_1}(t), C_{i,p_2}(t), \ldots\ldots, C_{i,p_N}(t)\right]^T$$

For the cases of modulated and unmodulated QEW, we get different $c_{p_n}^{(0)}$ as discussed in previous sections, which forms the initial vector elements with $C_{i,p_n}(t_0) = C_i(t_0)c_{p_n}^{(0)}$ at $t_0 = -\infty$. Summing all the scattering processes from $p_n$ to $p_m$, the differential equation can be expressed in a tensor form as:

$$\frac{d}{dt}v_i(t) = U^{ij}(t)v_j(t)$$

where $U^{ij}(t)$ determines the evolution of $v_i$, based on the state $v_j$, and its elements are

$$U_{nm}^{ij}(t) = \frac{\Delta p}{2\pi i\hbar^2}\widetilde{M}_{ij}(p_m-p_n)e^{-i(E_{p_n}-E_{p_m}-E_{ij})t/\hbar}.$$

With discretization of the time domain, the evolution of entangled free-electron and bound electron system is expressed as

$$\begin{pmatrix}v_1(t_{i+1})\\v_2(t_{i+1})\end{pmatrix} = \begin{pmatrix}1 & U^{12}(t_i)\Delta t\\ U^{21}(t_i)\Delta t & 1\end{pmatrix}\begin{pmatrix}v_1(t_i)\\v_2(t_i)\end{pmatrix}$$

with $\Delta t = t_{i+1} - t_i$. Finally, we get

$$\hat{\rho}(t_f) = \begin{pmatrix}v_1(t_f)\\v_2(t_f)\end{pmatrix}\begin{pmatrix}v_1^*(t_f) & v_2^*(t_f)\end{pmatrix}$$

4. N. F. Ramsey, A molecular beam resonance method with separated oscillating fields. *Phys. Rev.* **78**, 695 (1950).
5. T. C. Weinacht, J. Ahn, P. H. Bucksbaum, Controlling the shape of a quantum wavefunction. *Nature* **397**, 233-235 (1999).
6. C. H. Bennett, D. P. DiVincenzo, Quantum information and computation. *Nature* **404**, 247-255 (2000).
7. J. M. Kindem, A. Ruskuc, J. G. Bartholomew, J. Rochman, Y. Q. Huan, A. Faraon, Control and single-shot readout of an ion embedded in a nanophotonic cavity. *Nature* **580**, 201-204 (2020).
8. D. Leibfried, R. Blatt, C. Monroe, D. Wineland, Quantum dynamics of single trapped ions. *Rev. Mod. Phys.* **75**, 281–324 (2003).
9. N. H. Bonadeo, J. Erland, D. Gammon, D. Park, D. S. Katzer, D. G. Steel, Coherent optical control of the quantum state of a single quantum dot. *Science* **282**, 1473-1476 (1998).
10. L. Robledo, H. Bernien, I. Van Weperen, R. Hanson, Control and coherence of the optical transition of single nitrogen vacancy centers in diamond. *Phys. Rev. Lett.* **105**, 177403 (2010).
11. A. Gover, A. Yariv, Free-Electron-Bound-Electron Resonant Interaction, *Phys. Rev. Lett.* **124**, 064801 (2020).
12. Morimoto, Y., & Baum, P. (2018). Diffraction and microscopy with attosecond electron pulse trains. *Nature Physics*, *14*(3), 252-256.
13. J. Verbeeck, H. Tian, P. Schattschneider, Production and application of electron vortex beams. *Nature* **467**, 301-304 (2010).
14. N. Voloch-Bloch, Y. Lereah, Y. Lilach, A. Gover, A. Arie, Generation of electron Airy beams. *Nature* **494**, 331-335 (2013).
15. R. Shiloh, Y. Lereah, Y. Lilach, A. Arie, Sculpturing the electron wave-function using nanoscale phase masks. *Ultramicroscopy* **144**, 26-31 (2014).
16. P. Baum, Quantum dynamics of attosecond electron pulse compression. *J. Appl. Phys.* **122**, 223105 (2017).
17. Y. Morimoto, P. Baum, Single-cycle optical control of beam electrons. *Phys. Rev. Lett.* **125**, 193202(2020).
18. G. M. Vanacore, I. Madan, F. Carbone, Spatio-temporal shaping of a free-electron wave function via coherent light–electron interaction. *Riv. del Nuovo Cim.* 1-31 (2020).
19. A. Feist, K. E. Echternkamp, J. Schauss, S. V. Yalunin, S. Schfer, C. Ropers, Quantum coherent optical phase modulation in an ultrafast transmission electron microscope. *Nature.* **521**, 200-203 (2015).
20. K.E. Priebe, C. Rathje, S.V. Yalunin, T. Hohage, A. Feist, S. Schafer, C. Ropers, Attosecond electron pulse trains and quantum state reconstruction in ultrafast transmission electron microscopy. *Nat. Photonics* **11**, 793-797 (2017).
21. L. U. C. A. Piazza, T. T. A. Lummen, E. Quinonez, Y. Murooka, B. W. Reed, B. Barwick, F. Carbone, Simultaneous observation of the quantization and the interference pattern of a plasmonic near-field. *Nat. Commun.* **6**, 6407(2015).
22. M. Kozák, T. Eckstein, N. Schonenberger, P. Hommelhoff, Inelastic ponderomotive scattering of electrons at a high-intensity optical travelling wave in vacuum. *Nat. Phys*. **14**, 121(2018).
23. O. Reinhardt, I. Kaminer, Theory of shaping electron wavepackets with light. ACS Photonics 7, 2859-2870 (2020).
24. Yalunin, S. V., Feist, A., & Ropers, C. (2021). Tailored high-contrast attosecond electron pulses for coherent excitation and scattering. *Physical Review Research*, 3(3), L032036.
25. Kirchner, F. O., Gliserin, A., Krausz, F., & Baum, P. (2014). Laser streaking of free electrons at 25 keV. *Nature Photonics*, 8(1), 52-57.
26. Kozák, M., Schönenberger, N., & Hommelhoff, P. (2018). Ponderomotive generation and detection of attosecond free-electron pulse trains. *Physical review letters*, *120*(10), 103203.
27. Morimoto, Y., & Baum, P. (2018). Attosecond control of electron beams at dielectric and absorbing membranes. *Physical Review A*, *97*(3), 033815.
28. Ryabov, A., Thurner, J. W., Nabben, D., Tsarev, M. V., & Baum, P. (2020). Attosecond metrology in a continuous-beam transmission electron microscope. *Science advances*, *6*(46), eabb1393.
39